\documentclass[aps,prl,twocolumn,groupedaddress,longbibliography]{revtex4-1}

\usepackage{graphicx}
\usepackage{amsmath,amssymb}
\usepackage{mdwmath}
\usepackage{braket}

\begin{document}


\title{Comprehensive understanding of parity-time transitions in $\mathcal{PT}$ symmetric photonic crystals with an antiunitary group theory}


\author{Adam Mock}
\email[]{mock1ap@cmich.edu}
\affiliation{School of Engineering and Technology, Central Michigan University, ET 100, Mount Pleasant, MI 48859, USA \\ and
Science of Advanced Materials Program Central Michigan University, Mount Pleasant, MI 48859, USA}


\date{\today}

\begin{abstract}

Electromagnetic materials possessing parity-time symmetry have received significant attention since it was discovered that
the eigenmodes of these materials possess either real-frequency eigenvalues or the eigenfrequencies appear in complex-conjugate
pairs.  Interestingly, some eigenstates of these systems show thresholdless $\mathcal{PT}$ transitions to the complex-conjugate regime, some
exhibit a transition as a function of the degree of non-Hermiticity and some show no $\mathcal{PT}$ transition at all.  While previous work
has provided some insight on the nature of $\mathcal{PT}$ transitions, this work lays out a general and rigorous mathematical framework that is able to predict,
based on symmetry alone, whether an eigenmode will exhibit a thresholdless $\mathcal{PT}$ transition or no $\mathcal{PT}$ transition at all.  
Developed within the context of ferromagnetic solids, Heesh-Shubnikov group theory is an extension of classical group theory that is applicable to 
antiunitary operators.  This work illustrates the Heesh-Shubnikov approach by categorizing the modes of a two-dimensionally periodic 
photonic lattice that possesses $\mathcal{PT}$ symmetry.

\end{abstract}


\maketitle

\section{}
\subsection{}
\subsubsection{}

\section{I. Introduction}

Electromagnetic (EM) systems described by non-Hermitian wave equations but invariant under the combined operation of parity $\mathcal{P}$ and time-inversion $\mathcal{T}$ have been the
subject of intense investigation in recent years~\cite{elganainy2007, guo2009, mostafazadeh2009a, mostafazadeh2009b, ruter2010, 
longhi2010b, ctyroky2010, benisty2011,ge2012, hodeaei2014, longhi2014, chang2014, peng2014, phang2015,miao2016}.  
Such systems have been shown to possess either real-frequency eigenvalues or paired complex-conjugate
eigenvalues~\cite{bender1998, bender1999, bender2002}.  In the latter case, the modes separate into gain or loss modes depending on the sign of the complex part of the frequency.  
And the spatial field distribution of the gain and loss modes show a preferential overlap with the amplifying and absorbing portions of the geometry, respectively.
This work provides a theoretical foundation for understanding $\mathcal{PT}$ transitions in two-dimensionally periodic systems with $\mathcal{PT}$ symmetry.  The modes in these systems exhibit a rich variety of
behavior much of which lacks a rigorous mathematical basis~\cite{mock2016,cerjan2016}.  \\

Preliminary studies of $\mathcal{PT}$ symmetric EM systems concerned two identical parallel waveguides, one of which provided gain and the other was
absorbing~\cite{elganainy2007, guo2009, ruter2010}.  More complicated spatial arrangements were studied using optical fiber, fiber amplifiers
and fiber splitters/combiners~\cite{makris2008, regensburger2012, regensburger2014}.  And fascinating recent work has exploited $\mathcal{PT}$ symmetry for laser mode 
control, optical isolation and vortex beam generation in ring resonator devices~\cite{hodeaei2014, feng2014, longhi2014, chang2014, peng2014, phang2015}.
However, periodic systems (especially with periodicity in dimensions higher than one) with $\mathcal{PT}$ symmetry have only begun to be studied~\cite{musslimani2008, makris2008, zhou2010,  lin2011, szameit2011, 
regensburger2012, feng2013, regensburger2014, zhu2014, alaeian2014, xie2014, yannopapas2014, kulishov2014, ge2015, wang2015,  zhu2015, agarwal2015,  
ge2015b, ding2015, turdeuev2015, cerjan2016, cerjan2016b, mock2016}.  This work shows how a group theory approach originally developed to classify the irreducible representations of
ferromagnetic solids can be used to predict $\mathcal{PT}$ transitions and degeneracies in the photonic band diagram.  \\

Materials whose optical properties are periodically modulated in two dimensions (also known as two-dimensional photonic crystals) are critical in numerous areas of 
optics including integrated photonics, 
photonic metasurfaces, optical sensing, and studies of quantum electrodynamics and embedded eigenvalues.  In each of these areas, understanding the nature of the electromagnetic modes
is a critical first step.  In relatively simple structures, modes can be characterized according to their parity about some axis of the system, and this information is typically
sufficient to estimate critical design metrics such as coupling and radiation efficiencies.  However, in two-dimensional (2D) photonic crystals (PCs), modes exhibit a richer variety of
behavior~\cite{mock2016}, and a more rigorous mode classification scheme is required.  Group theory has been extremely useful in the classification of modes in 2D PCs~\cite{sakoda2001,srinivasan2002}.  
Additionally, group theory can predict mode degeneracies.  However, application of ``classical'' group theory to $\mathcal{PT}$ symmetric systems is hampered by the fact that the 
time-inversion operation is antiunitary, and classical representation theory assumes all operations can be represented by unitary matrices (or matrices that are equivalent under
unitary transformations).  Fortunately, Heesh-Shubnikov group
theory was developed to handle such situations with its associated corepresentations.  Heesh-Shubnikov group theory was developed to analyze magnetic ordering in 
ferromagnetic materials in the middle decades of the twentieth century, and here we show that it continues to be instrumental in the understanding of $\mathcal{PT}$ symmetric photonics.  \\

\begin{figure}
\centering
\includegraphics[width=7.0cm]{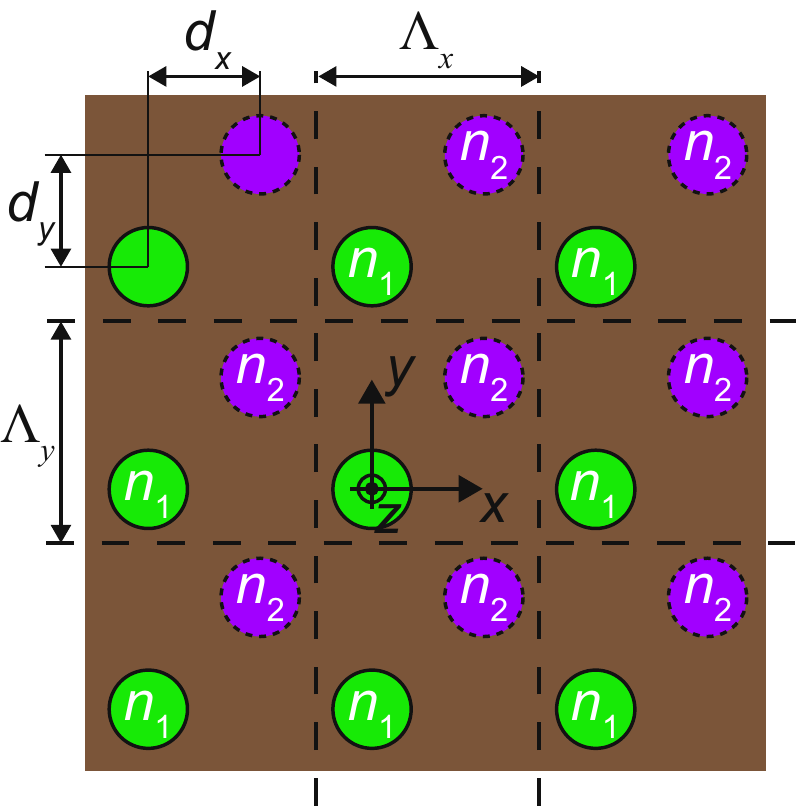}%
\caption{Schematic diagram showing the two-dimensional $\mathcal{PT}$ symmetric photonic lattice.  Regions labeled $n_1$ provide gain ($n_1 = n_r + i n_i$),
and regions labeled $n_2$ provide loss ($n_2 = n_r - i n_i$) for positive $n_r$ and $n_i$.   \label{2dGeom} }
\end{figure}

\section{II. Heesh-Shubnikov Group Theory}

Formally, $\mathcal{PT}$ symmetry concerns electromagnetic systems whose refractive index obeys $\mathcal{P} n(\mathbf{r}) = n(\mathbf{r})^*$ where the parity operator $\mathcal{P}$ inverts
one or more spatial coordinates.  However, the interesting consequences of $\mathcal{PT}$ symmetry hold even when the parity operator is generalized to be any spatial symmetry operation while keeping
the total amount of gain and loss balanced.  
In this paper, we concern ourselves with the geometry depicted in Fig.~\ref{2dGeom} with $d_x = d_y = \Lambda/2$ where $\Lambda_x=\Lambda_y=\Lambda$ is the lattice constant
for a square lattice.  With the origin centered on a gain rod, it is apparent that the structure is invariant under point symmetry operations in the $C_{4v} (4mm)$ point group.  
Further, the lattice is invariant under the combination of a spatial shift of $\boldsymbol{\tau} = \frac{\Lambda}{2}(\boldsymbol{x} + \boldsymbol{y})$ and complex-conjugation (or $\mathcal{T}$).  
In this sense the generalized $\mathcal{PT}$ invariance is $\{E | \frac{\Lambda}{2}(\boldsymbol{x} + \boldsymbol{y}) \} n(x,y) = n(x,y)^*$ where $E$ is the identity operator and the Seitz
notation is used to represent the combined point symmetry operation ($E$) and spatial shift ($\boldsymbol{\tau}$).  After the lattice has been shifted and complex conjugated,
one can again apply the point symmetry operations in $C_{4v}$ and the geometry will remain invariant.  Ultimately, the lattice possesses two $C_{4v}$ point group symmetry
centers centered on the gain and loss rods.  The full space group of the geometry is given by $\mathcal{M} = C_{4v} + \{ \mathcal{T} | \frac{\Lambda}{2}\frac{\Lambda}{2}  \} C_{4v}$
where $C_{4v}$ represents the set of symmetry operations in the group $C_{4v}$ [that is, let $C_{4v} = (E, C_2, C_4, C^{-1}_4, \sigma_x, \sigma_y, \sigma_d, \sigma_d')$] and 
$\{ \mathcal{T} | \frac{\Lambda}{2}\frac{\Lambda}{2}\} \equiv \{\mathcal{T} | \frac{\Lambda}{2}(\boldsymbol{x} + \boldsymbol{y}) \}  $.

The central question this paper addresses is whether thresholdless $\mathcal{PT}$ transitions can be predicted from symmetry.  Group theory has been instrumental in understanding the qualitative
features of both electronic and photonic band structures.  Notably, it allows the determination of band degeneracies at both high and low symmetry points in reciprocal space from
the irreducible representations of the groups.
The presence of time-inversion introduces complex-conjugation into the symmetry operations considered here which makes the operators anti-unitary.   Because the irreducible representations of 
``classical'' groups (those without complex-conjugation) assume the symmetry operators are unitary, an alternative approach for constructing the irreducible representations is needed.
While the Heesh-Shubnikov group theory was pioneered by their namesake, their exposition by Wigner played a significant role in their application to practical 
problems~\cite{heesh1929,shubnikov1951, shubnikov1964,wigner1959, cracknell1975a, cracknell1975b}.  
Heesh-Shubnikov groups are also referred to as a magnetic groups or black and white groups. In the context of periodic structures they apply to crystals whose lattice sites have a property
that can take on one of two values.

The Heesh-Shubnikov group theory apparatus actually provides two critical pieces of information.  In addition to providing the irreducible representations (more commonly referred to as ``corepresentations'') 
of the anti-unitary operators, it also
provides a categorization system that can be used to determine whether thresholdless $\mathcal{PT}$ transitions will occur.  
The categorization is based on a sum-rule test developed by Dimmock and Wheeler~\cite{frobenius1906,dimmock1962}.
They discovered that corepresentations of $\mathcal{M}$ fall into three 
categories~\cite{wigner1959,ElBatanouny2008}, and the categorization is determined by summing the classical characters of the elements resulting from squaring
all symmetry elements that include the antiunitary operator $\mathcal{T}$:
\begin{equation}
  \sum_{B \in \mathcal{W}} \chi(B^2) = \begin{cases}
                                                                     n & \textrm{Type (a),}  \\
                                                                    -n & \textrm{Type (b),}    \\
                                                                    0   &  \textrm{Type (c),} 
  \end{cases} \label{dmT}
\end{equation}
where $\chi(R)$ is the character of the classical representation of $R$, $n$ is the order of the unitary subgroup (to be explained below) and $\mathcal{W}$ is the set of
antiunitary operators (those containing $\mathcal{T}$).  In previous work, we discovered that Type (a) corepresentations correspond to a single representation of the unitary subgroup, 
and no new $\mathcal{PT}$ degeneracy is introduced.  The eigenfrequencies of Type (a) modes remain real.  Type (c) corepresentations contain two inequivalent representations of 
the unitary subgroup, and new $\mathcal{PT}$ degeneracy is introduced~\cite{wigner1959,ElBatanouny2008,mock2016b}.  Modes with Type (c) corepresentations exhibit thresholdless $\mathcal{PT}$
symmetry transitions, so their eigenfrequencies come in complex-conjugate pairs even in the limit of inifinitesimal non-Hermiticity (that is, infinitesimal amounts of gain and loss).
Type (b) corepresentations contain the same single representation of the unitary subgroup twice, and new $\mathcal{PT}$ degeneracy may appear.

Returning to the space group $\mathcal{M}$ defined above, we note that the $C_{4v}$ part forms a unitary subgroup of index 2, and the antiunitary elements
form a coset of $C_{4v}$.  Generally, $\mathcal{PT}$ symmetric geometries will have symmetry groups of the form $\mathcal{M} = \mathcal{N} + A \mathcal{N}$ where $\mathcal{N}$
is a unitary subgroup, and $A$ is an antiunitary symmetry operator.  Cracknell classifies Heesh-Shubnikov groups of this form as Type IV~\cite{cracknell1975a,cracknell1975b}.

Type (a) corepresentations maintain the dimensionality of the irreducible representations of the corresponding unitary subgroup $\mathcal{N}$.  The dimensionalities of Type (b) and (c)
corepresentations are twice that of the corresponding unitary subgroup.  This is consistent with the observation that eigenstates with complex-conjugate eigenfrequencies
associated with thresholdless $\mathcal{PT}$ transitions transform into each other under antiunitary symmetry operations, whereas those with real frequencies transform only into themselves.

\begin{figure}
\centering
\includegraphics[width=8.6cm]{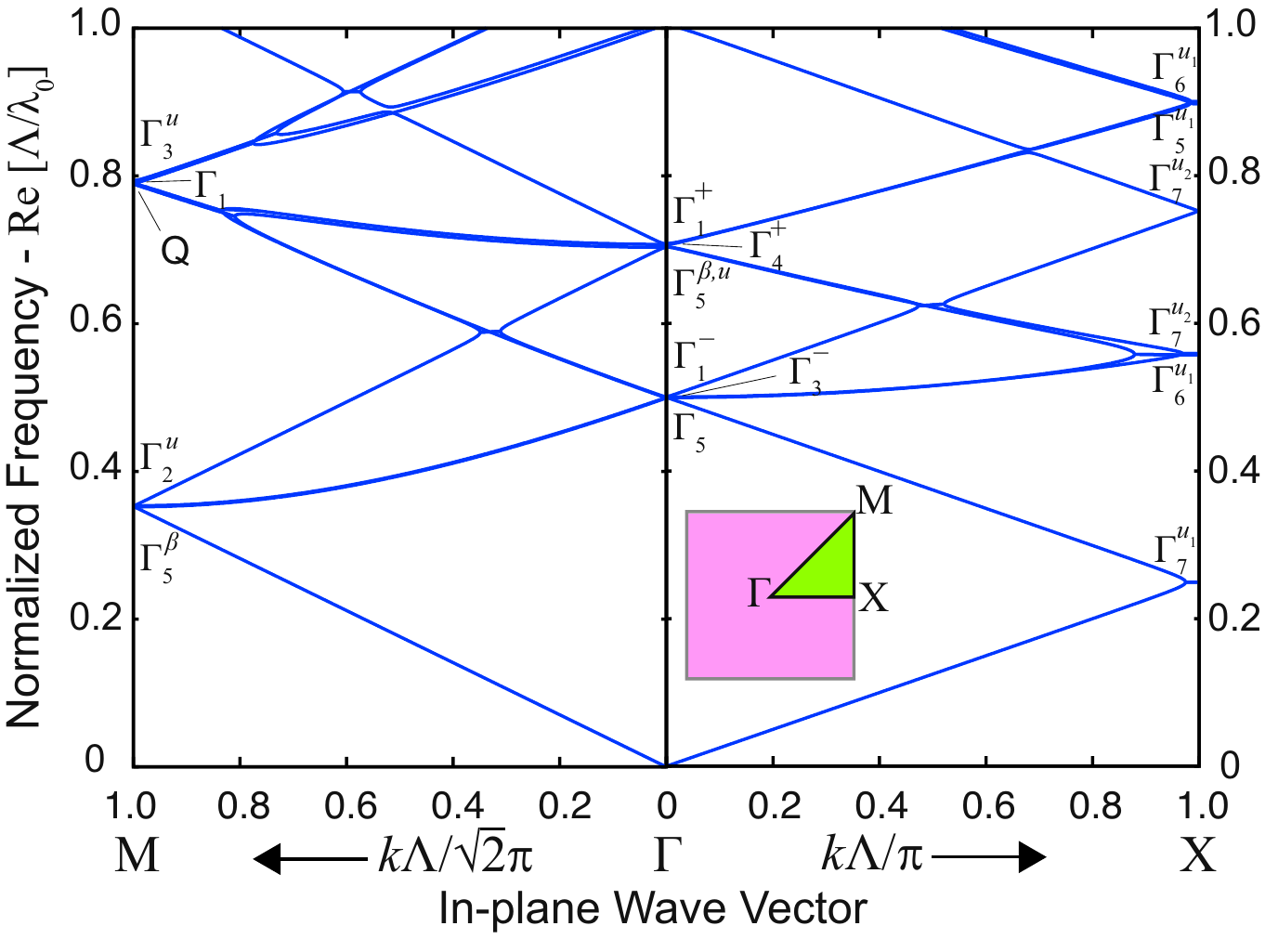}%
\caption{Photonic band structure for lattice shown in Fig.~\ref{2dGeom} calculated using the plane wave expansion method~\cite{plihal1991,sakoda2001}. 
Results are shown for the TE polarization $\{E_x, E_y, H_z\}$.  Labels on the bands refer to specific corepresentations listed in Ref.~\cite{smprx3}.
Inset depicts the first Brillouin zone with labels at the high symmetry points.   \label{bs} }
\end{figure}

In the following we illustrate these concepts at high-symmetry
points in the photonic band structure shown in Fig.~\ref{bs}. Immediately one sees a variety of interesting behavior including thresholdless 
$\mathcal{PT}$ symmetry transitions at the high-symmetry X and M points.
Furthermore, the structure also supports ``classical'' degeneracy or points where bands coalesce but do not yield a $\mathcal{PT}$ symmetry transition (eigenfrequencies remain real).
This paper provides a general framework that explains where in the photonic band diagram $\mathcal{PT}$ transitions are expected to occur.  Our analysis will use only the 
TE polarization $\{E_x, E_y, H_z\}$ to illustrate the concept.  The approach applies to the TM polarization as well.

\section{III. Results}

Here we how the Heesh-Shubnikov group theory approach can predict the nature of $\mathcal{PT}$ transitions at the high symmetry points $\Gamma$ [$\mathbf{k} = 0$],
X [$\mathbf{k} = \frac{\pi}{\Lambda} \boldsymbol{x}$] and M [$\mathbf{k} = \frac{\pi}{\Lambda} (\boldsymbol{x} + \boldsymbol{y})$].  In each case the points of
interest will be band crossings occurring in the empty lattice (a lattice with infinitesimally small periodic perturbation)~\cite{ge2014}.  What we find ultimately is that every empty lattice
band crossing exhibits either a thresholdless $\mathcal{PT}$ transition or no $\mathcal{PT}$ transition at all.  The group theory approach described here shows how symmetry can be used
to distinguish and predict which points will or will not exhibit a $\mathcal{PT}$ transition.

\subsection{III.A Absence of $\mathcal{PT}$ transitions at the $\Gamma$ point [$\mathbf{k} = 0$]} 

Symmetry analysis in periodic structures is facilitated by identifying the group of $\mathbf{k}$ or \textit{little group}.
In classical group theory, the little group consists of elements of the full space group which send a particular
$\mathbf{k}$ into $\mathbf{k} + \mathbf{K}$ where $\mathbf{K}$ is a reciprocal lattice vector~\cite{sakoda2001,tinkham1964}. 
However, for \textit{Heesh-Shubnikov little groups} (HSLG) that 
include antiunitary operators, such a group includes (i) unitary elements of the space group that send $\mathbf{k}$ into $\mathbf{k} + \mathbf{K}$ (as before) and 
(ii) antiunitary elements of the space group that send $\mathbf{k}$ into $-\mathbf{k} + \mathbf{K}$~\cite{cracknell1975b}.

\begin{figure}
\centering
\includegraphics[width=8.6cm]{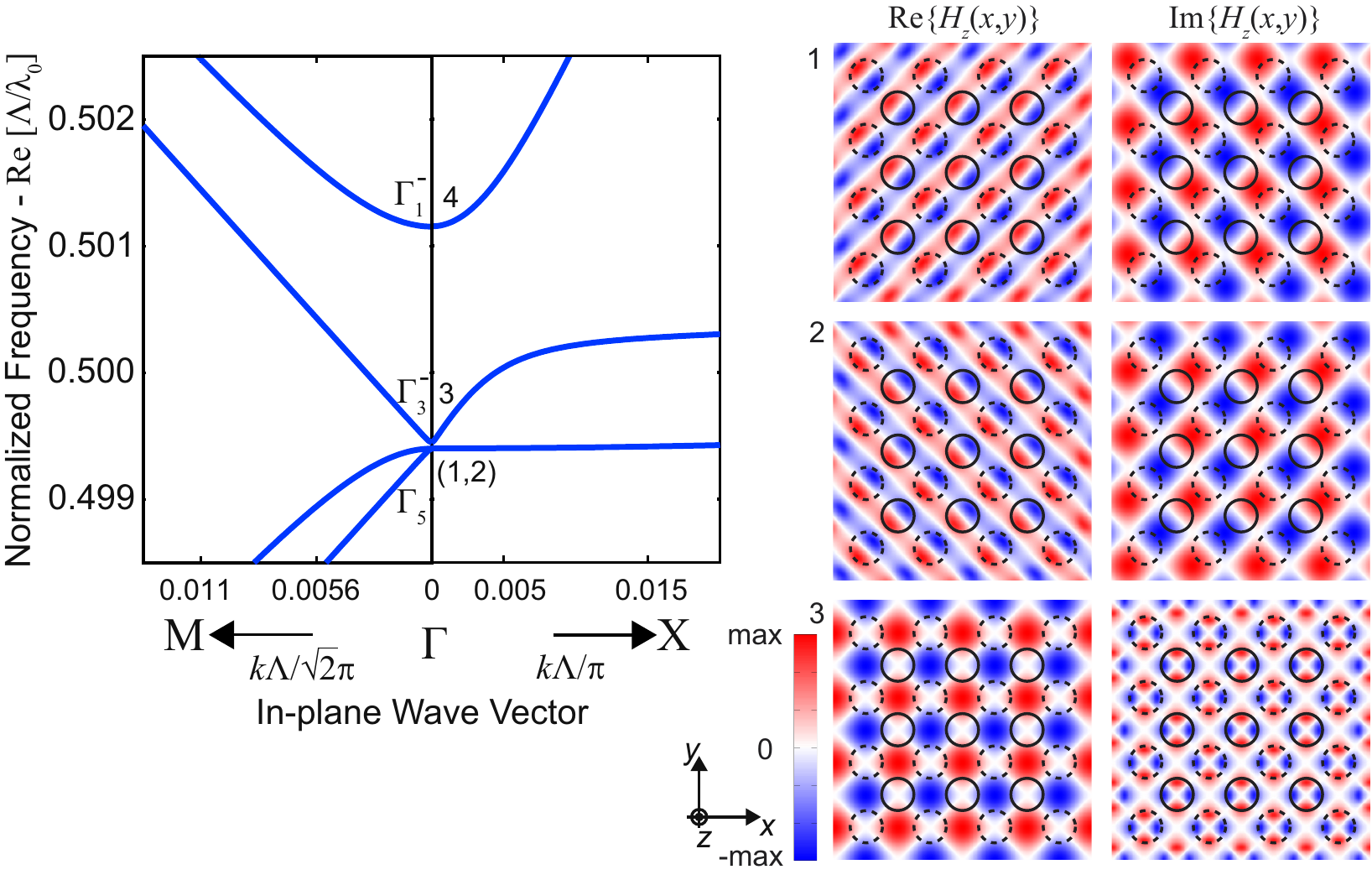}%
\caption{Left side shows a detailed view of the bandstructure at the $\Gamma$ point near the frequency $\Lambda/\lambda_0 = 0.5$.
Bands labeled (1,2) are degenerate at $\mathbf{k} = 0$.  The band labeled 3 crosses $\mathbf{k} = 0$ at a frequency slightly above the (1,2) bands and is not
degenerate with them.  Right side shows $H_z(x,y)$ for the modes labeled 1, 2 and 3.  For modes 1 and 2, the amplitude scale of the real part is eight times smaller than
that of the imaginary part.  For mode 3, the amplitude scale of the imaginary part is twenty times smaller than that of the real part.
 \label{k0zm} }
\end{figure}

Based on the Bloch form for modes of periodic systems, one can 
use a representation of the space group, $\exp[i \mathbf{k} \cdot (m\boldsymbol{x} + n\boldsymbol{y})\Lambda]$, where $\{m,n\} \in \mathbb{Z}$~\cite{heine1960, mock2010c}.  
At $\mathbf{k} = 0$ (the $\Gamma$ point), the analysis is simplified since the space group representation $\exp[i \mathbf{k} \cdot (m \boldsymbol{x} + n \boldsymbol{y})] = 1$ for all $m$ and $n$.  
In this case, the HSLG is the same as the full space group $\mathcal{M}^{\mathbf{k} = 0} = \mathcal{M} = C_{4v} + \{ \mathcal{T} | \frac{\Lambda}{2}\frac{\Lambda}{2}  \} C_{4v}$.  
More details on the specific symmetry operators are provided in Ref.~\cite{smprx3}.

Squaring each element in the group $\{ \mathcal{T} | \frac{\Lambda}{2}\frac{\Lambda}{2}  \} C_{4v}$ yields the elements $(E, E, E, E, E, E, C_2, C_2 )$.  Summing the characters of these elements according to 
the Dimmock-Wheeler test yields $\sum_{B \in \mathcal{W}} \chi(B^2) = 8$ for all of the classical representations of $C_{4v}$.  Therefore,
the modes of the $\mathcal{PT}$ symmetry lattice shown in Fig.~\ref{2dGeom} at $\mathbf{k} = 0$ are of Type (a), and no thresholdless $\mathcal{PT}$ transitions are expected there.
The group $C_{4v}$ contains four one-dimensional (1D) representations and one two-dimensional (2D) representation, so one anticipates non-degenerate and 2D degenerate modes at $\mathbf{k} = 0$.  

The components of the $i$th Type (a) corepresentation $\boldsymbol{\Gamma}_i$ for the unitary elements $R \in \mathcal{N}$ are given by 
$\boldsymbol{\Gamma}_i(R) = \boldsymbol{\Delta}_i(R)$ where $\boldsymbol{\Delta}_i(R)$ is the $i$th classical representation of $R$ in $\mathcal{N}$.
The components of the $i$th Type (a) corepresentation $\boldsymbol{\Gamma}_i$ for the antiunitary elements $R \in \mathcal{W}$ are given by
$\boldsymbol{\Gamma}_i(RA) = \boldsymbol{\beta} \boldsymbol{\Delta}_i(R)$ where $A \in \mathcal{W}$ is an arbitrary but fixed antiunitary operator and $\boldsymbol{\beta}$ is a matrix that satisfies
$\boldsymbol{\beta} \boldsymbol{\beta}^* = \boldsymbol{\Delta}_i(A^2)$~\cite{wigner1959, ElBatanouny2008}.  A full corepresentation table which includes the labeling
scheme in Fig.~\ref{bs} and Fig.~\ref{k0zm}(a) is provided in Ref.~\cite{smprx3}.

Fig.~\ref{k0zm}(a) shows a detailed version of the photonic bandstructure shown in Fig.~\ref{bs} at the $\Gamma$ point around the frequency $\Lambda/\lambda_0 = 0.5$.  
One sees a 2D degenerate set of modes (1,2) and non degenerate
modes (3) and (4).  However, all eigenfrequencies are real, and no $\mathcal{PT}$ transitions occur.  Considering the field distributions $H_z(x,y)$ depicted in Fig.~\ref{k0zm}(b), 
one sees that the 2D degenerate modes transform either into themselves or into each other under the various symmetry operations.  The non-degenerate mode transforms into itself
only (with corresponding character).  Regardless of degeneracy, strictly real eigenfrequencies at $\mathbf{k}=0$ is consistent with the equal energy overlap of the 
eigenmodes with the gain and loss rods.  This is also consistent with the previous observation that the energy distribution of modes at $\mathbf{k}=0$ has 
a spatial periodicity of $\Lambda/2$~\cite{mock2016,mock2016b}.

\begin{figure}
\centering
\includegraphics[width=8.6cm]{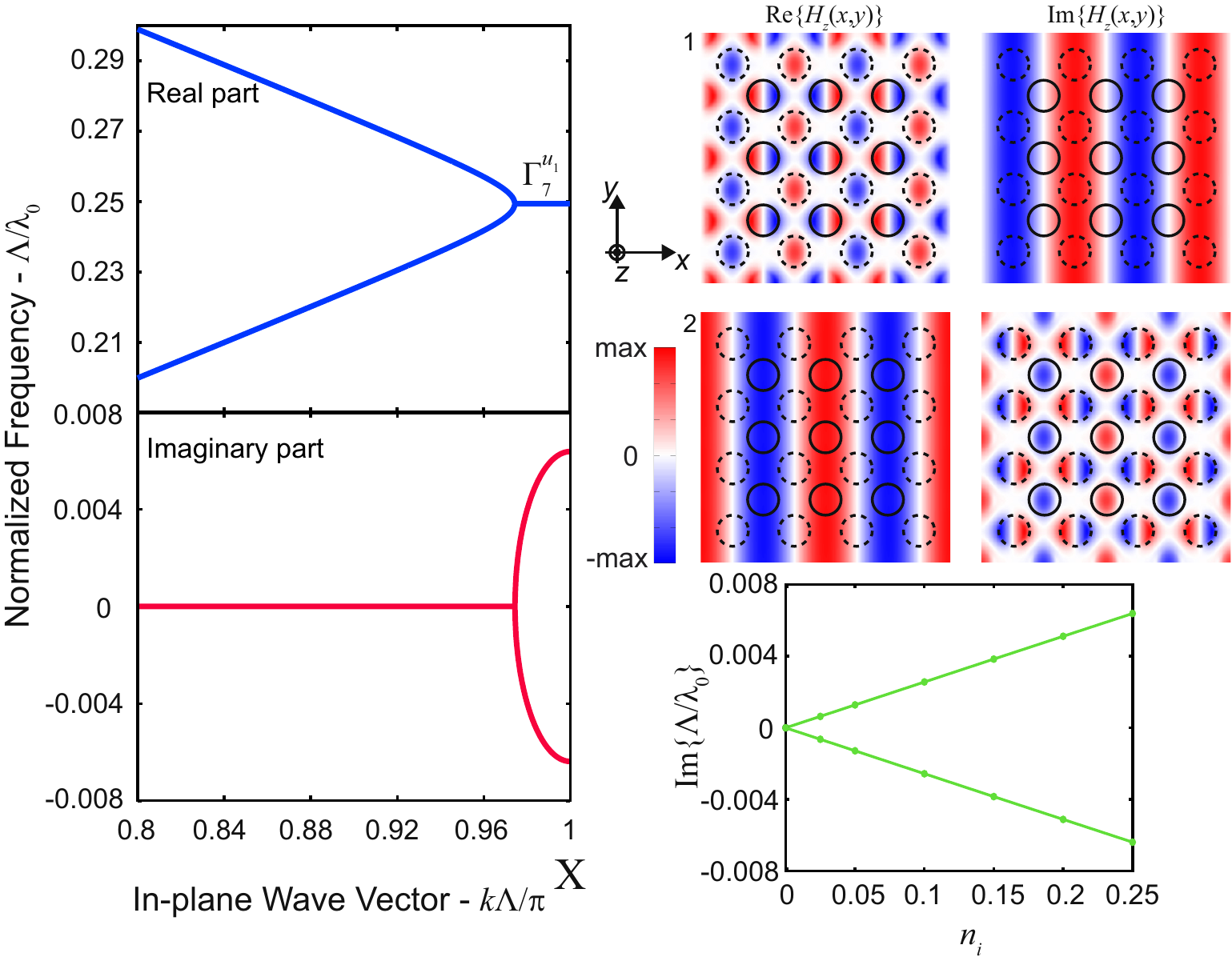}%
\caption{Left side shows a detailed view of the bandstructure at the X point near the (real) frequency $\Lambda/\lambda_0 = 0.25$.
Right side shows $H_z(x,y)$ for the gain and loss modes associated with the $\mathcal{PT}$ transition at the X point.  
For mode 1 (2), the amplitude scale of the real (imaginary) part is 0.053 times that of the imaginary (real) part.  
Bottom right figure depicts the imaginary part of the frequency as a function of the imaginary part of the refractive index.  The trend is characteristic of a thresholdless
$\mathcal{PT}$ transition.
\label{kXzm} }
\end{figure}

\subsection{III.B Thresholdless $\mathcal{PT}$ transitions at the X point [$\mathbf{k} = \frac{\pi}{\Lambda} \boldsymbol{x}$]} 

\begin{figure}
\centering
\includegraphics[width=8.6cm]{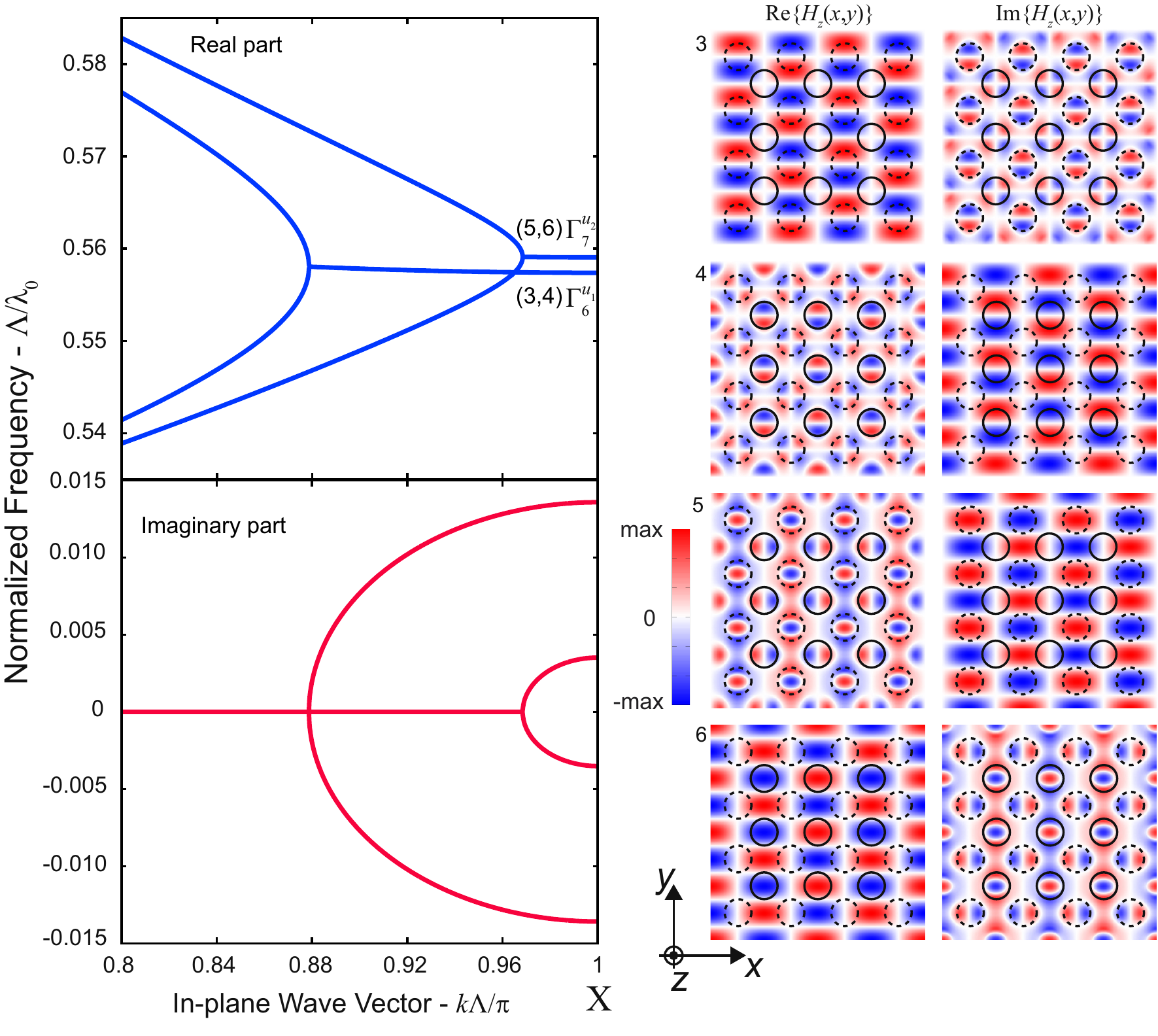}%
\caption{Left side shows a detailed view of the bandstructure at the X point near the (real) frequency $\Lambda/\lambda_0 = 0.56$ where two sets of complex-conjugate pairs
result from $\mathcal{PT}$ transitions.
Right side shows the corresponding $H_z(x,y)$.  
For modes 1 and 4, the amplitude scale of the imaginary part is ten times smaller than that of the real part.
For modes 2 and 3, the amplitude scale of the real part is ten times smaller than that of the imaginary part.
 \label{kXzm2} }
\end{figure}

The symmetry elements in the HSLG at $\mathbf{k} = \frac{\pi}{\Lambda} \boldsymbol{x}$ include $(E, C_2, \sigma_x, \sigma_y)$ and
$ \{ \mathcal{T} | \frac{\Lambda}{2},\frac{\Lambda}{2}  \} (E, C_2, \sigma_x, \sigma_y)$.  
Because $\exp[i \mathbf{k} \cdot (\boldsymbol{x}m + \boldsymbol{y}n)\Lambda)] = \exp[i \pi m]$ which is 1 for $m$ even and
$-1$ for $m$ odd, these two space group representations must be explicitly included in the group.  
This doubles the size of the HSLG, and the resulting unitary subgroup is isomorphic to $D_{2h} (mmm)$.
More details on the symmetry operators are provided in Ref.~\cite{smprx3}.
Ultimately the HSLG at $\mathbf{k} = \boldsymbol{x} \frac{\pi}{\Lambda}$ is $\mathcal{M}^{\mathbf{k} = \boldsymbol{x} \frac{\pi}{\Lambda}} = D_{2h} + \{ \mathcal{T} | \frac{\Lambda}{2}\frac{\Lambda}{2} \} D_{2h} $.
The group $D_{2h}$ has eight 1D representations.  Performing the Dimmock-Wheeler test yields four Type (a) and four Type (c) corepresentations.
Enforcing the physical constraint that the character of $\exp[i \pi m]$ switch sign for even and odd $m$, leaves only the four Type (c) corepresentations. 
Therefore, we anticipate that thresholdless $\mathcal{PT}$ transitions are expected at \textit{every} empty-lattice band crossing at $\mathbf{k} = \boldsymbol{x} \frac{\pi}{\Lambda}$.  
In this case only pairwise coupled modes with complex-conjugate frequencies are expected; no higher order degeneracy (or
nondegeneracy) is anticipated.  Figures~\ref{kXzm} and~\ref{kXzm2} show detailed views of the lowest three bands at the X point.  Indeed in each of these
cases a $\mathcal{PT}$ transition is seen near $\mathbf{k} = \frac{\pi}{\Lambda}\boldsymbol{x}$.  Moreover, the bottom right panel of Fig.~\ref{kXzm} shows that the exceptional point approaches
$\mathbf{k} = \frac{\pi}{\Lambda}\boldsymbol{x}$ as $n_i \rightarrow 0$ indicating that the $\mathcal{PT}$ transition is thresholdless.  \\

The components of the $i$th Type (c) corepresentation $\boldsymbol{\Gamma}_i$ for the unitary elements 
 $R \in \mathcal{N}$ are given by~\cite{wigner1959, ElBatanouny2008}
\begin{equation}
\boldsymbol{\Gamma}_i(R) = \begin{pmatrix}
    \boldsymbol{\Delta}(R)  &  \mathbf{0} \\
    \mathbf{0}               &   \boldsymbol{\Delta}^*(S^{-1}RS) 
\end{pmatrix}
\label{eqcorepU}
\end{equation}
where $A = S\mathcal{T}$.  The components of the $i$th Type (c) correpresentation $\boldsymbol{\Gamma}_i$ for the antiunitary elements $R \in \mathcal{W}$ are given by
\begin{equation}
\boldsymbol{\Gamma}_i(R) = \begin{pmatrix}
   \mathbf{0}                    &  \boldsymbol{\Delta}^*(A^{-1}R) \\
   \boldsymbol{\Delta}(RA)   &  \mathbf{0}  
\end{pmatrix}
\label{eqcorepAU}
\end{equation}
where these matrices operate on the loss/gain mode pair 
\begin{equation}
H = \begin{pmatrix}
H_l \\
H_g
\end{pmatrix}
\end{equation}
where $l$ ($g$) refers to loss (gain).

These equations show precisely how two-dimensional corepresentations representing the coupled complex-conjugate eigenfrequency pairs are constructed 
from nominally 1D irreducible representations of the unitary subgroup.  Further, one sees that the corepresentations of the unitary operators are diagonal, so that they
simply transform the modes making up the complex-conjugate pair into themselves.  In contrast, the corepresentations of the antiunitary operators are antidiagonal
which means the antiunitary operators transform these modes into each other.  A critical aspect of the antiunitary operators is the shift by $(\boldsymbol{x} + \boldsymbol{y})\frac{\Lambda}{2}$.
Almost always, the gain mode and loss mode can be transformed into each other (with an appropriate constant multiplier) with this shift.  This can be seen
in the field profiles displayed in Figs.~\ref{kXzm} and~\ref{kXzm2} where there is an additional factor of $\pm i$ involved in the shift.  The specific transformation matrices
are provided in tables in Ref.~\cite{smprx3}.

\begin{figure}
\centering
\includegraphics[width=8.6cm]{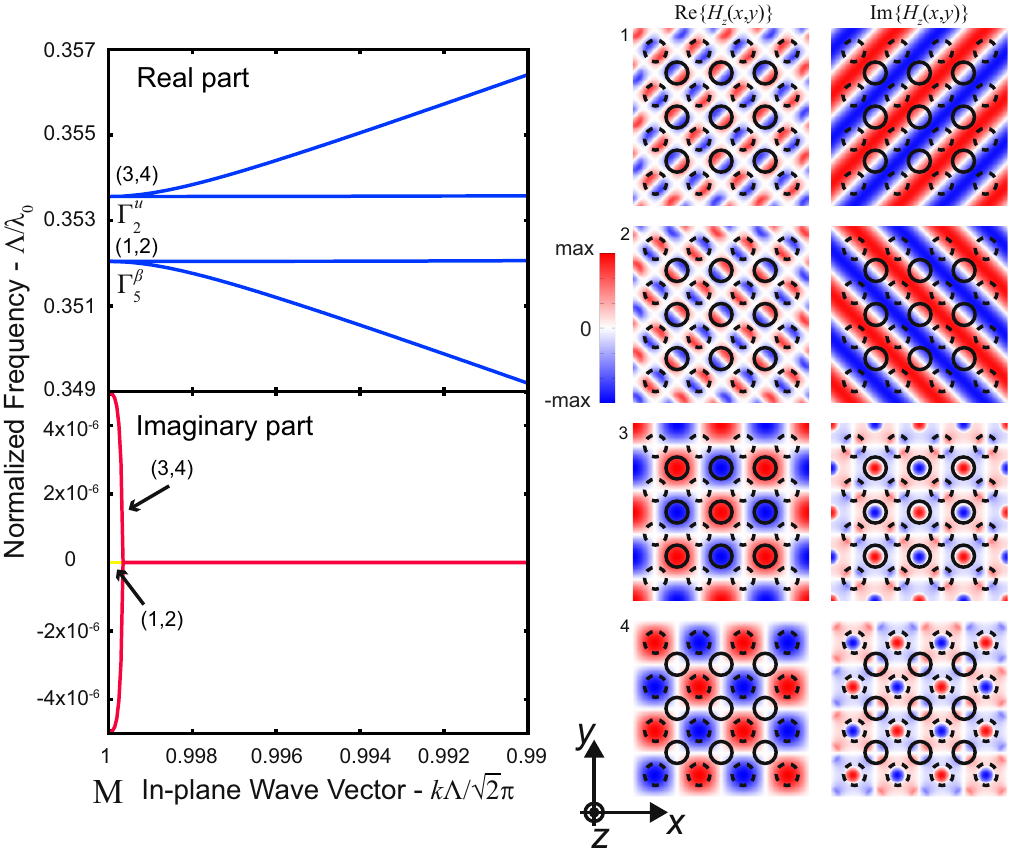}%
\caption{Left side shows a detailed view of the bandstructure at the M point near the (real) frequency $\Lambda/\lambda_0 = 0.353$ where one set of modes exhibits a $\mathcal{PT}$ 
transition, and another exhibits two-dimensional ``classical'' degeneracy.
Right side shows the corresponding $H_z(x,y)$.  
For modes 1 and 2, the amplitude scale of the real part is twenty times smaller than that of the imaginary part.
For modes 3 and 4, the amplitude scale of the imaginary part is 12.5 times smaller than that of the real part.
\label{kMzm1} }
\end{figure}

\subsection{III.C Behavior at the M point [$\mathbf{k} = (\boldsymbol{x} + \boldsymbol{y}) \frac{\pi}{\Lambda}$] } 

Lastly, consider $\mathbf{k} = (\boldsymbol{x} + \boldsymbol{y}) \frac{\pi}{\Lambda}$ which is the M point.  
At the M point, the applicable point symmetry operations include all of those in $C_{4v}$.   Because $\exp[i \mathbf{k} \cdot (\boldsymbol{x}m + \boldsymbol{y}n)\Lambda)] = \exp[i \pi (m+n)]$ which 
is 1 for $m+n$ even and $-1$ for $m+n$ odd, these two space group representations must be explicitly included in the group.  
This doubles the size of the HSLG, and the resulting 16-element unitary subgroup is isomorphic to $D_{4h} (4/mmm)$.
Altogether the HSLG at $\mathbf{k} = (\boldsymbol{x} + \boldsymbol{y}) \frac{\pi}{\Lambda}$ is 
$\mathcal{M}^{\mathbf{k} = (\boldsymbol{x} + \boldsymbol{y}) \frac{\pi}{\Lambda}} = D_{4h} + \{ \mathcal{T} | \frac{\Lambda}{2}\frac{\Lambda}{2} \} D_{4h} $.  
More details on the particular symmetry operations are provided in Ref.~\cite{smprx3}.  The group $D_{4h}$ contains four applicable 1D Type (c) representations
that produce two-dimensional corepresentations associated with thresholdless $\mathcal{PT}$ transitions.  It also contains one applicable 2D Type (a) representation, so one
expects a doubly degenerate mode at M in the non-Hermitian system but with real eigenvalues.  There are five other irreducible representations of $D_{4h}$ but these
are deemed not applicable because they do not change sign between elements with even and odd values of $m+n$.

Figure~\ref{kMzm1} shows a detailed view of the band structure in Fig.~\ref{bs} at the M point near the real frequency $\Lambda/\lambda_0 = 0.35$.
Two sets of bands converge toward degeneracy.  The bottom set belong to the Type (a) ``classically" doubly degenerate corepresentation, and the imaginary part
of their eigenfrequency remains zero all the way to $\mathbf{k} =  (\boldsymbol{x}+\boldsymbol{y})\frac{\pi}{\Lambda}$.  The top set belong to a Type (c) corepresentation, and they
exhibit a $\mathcal{PT}$ transition at around $\mathbf{k} = 0.9995  (\boldsymbol{x}+\boldsymbol{y})\frac{\pi}{\Lambda}$.  The spatial field distributions exhibit expected behavior.  The classically
degenerate modes transform either into themselves or into each other maintaining equal energy overlap with the gain and loss rods.  The modes exhibiting a $\mathcal{PT}$
transition separate into gain and loss modes, and the antiunitary symmetry operators transform these two modes into each other.

\begin{figure}
\centering
\includegraphics[width=8.6cm]{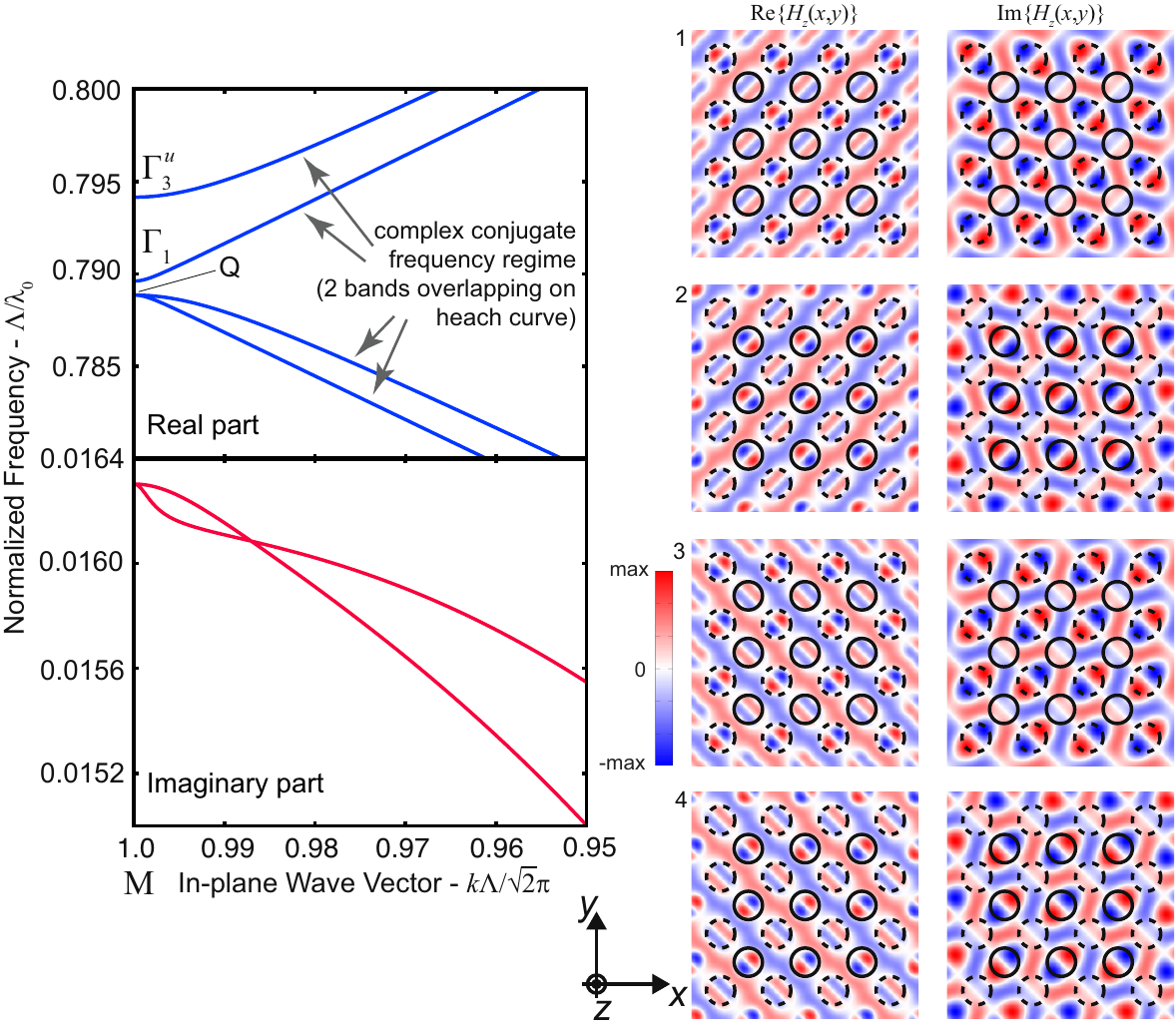}%
\caption{Left side shows a detailed view of the bandstructure at the M point near the (real) frequency $\Lambda/\lambda_0 = 0.79$.  The curves shown in the imaginary 
part correspond to the bottom two curves on the real part.
Right side shows the corresponding $H_z(x,y)$.  
The amplitude scale of the real part is 10 times smaller than that of the imaginary part.
 \label{kMzm2} }
\end{figure}

\subsection{III.D Unexpected Behavior at the M point [$\mathbf{k} = (\boldsymbol{x} + \boldsymbol{y}) \frac{\pi}{\Lambda}$] } 

We show in Figure~\ref{kMzm2} another detailed view of the M point, this time near the real frequency $\Lambda/\lambda_0 = 0.79$.  The empty lattice bandstructure possesses eight-fold degeneracy
at this point.  When the $\mathcal{PT}$ symmetric lattice is introduced, the eight-fold degeneracy is lifted.  The view afforded by Fig.~\ref{kMzm2} depicts 4 bands all of which are in the
complex-conjugate frequency regime, so each band consists of two overlapping bands with the same real frequency.  The $\mathcal{PT}$ transition point can be seen in Fig.~\ref{bs} and
occurs around $k\Lambda/(\sqrt{2}\pi) = 0.8$.  

This point is highlighted because as these four bands approach the M point, the bottom two coalesce resulting in a doubly degenerate
set of modes in the complex-conjugate frequency regime (point labeled Q).  That is, the four modes have two sets of equal complex-conjugate frequencies.
This is a surprising observation since the Heesh-Shubnikov analysis predicts either non-degenerate Type (c) representations (two coupled modes with 
complex-conjugate frequencies) or doubly degenerate Type (a) representations that remain in the real-frequency domain.  The lowest frequency modes shown in Fig.~\ref{kMzm2}
appear to exhibit properties of a doubly degenerate Type (c) representation even though one does not exist in the present analysis.

\begin{figure}
\centering
\includegraphics[width=8.6cm]{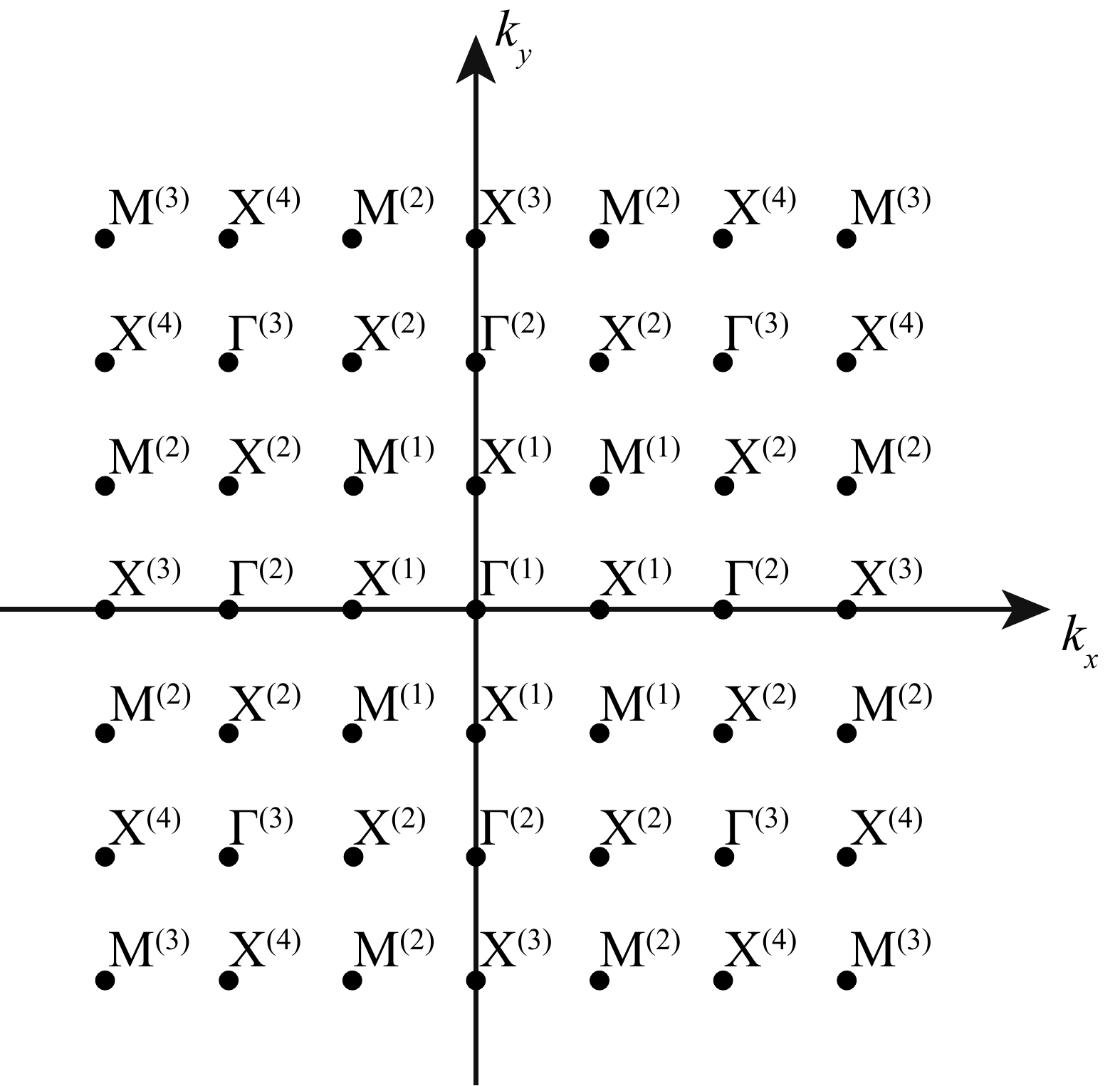}%
\caption{Reciprocal lattice diagram showing equivalent high-symmetry points connected by reciprocal lattice vectors.  The superscripts rank their distance from the origin. \label{rlmap} }
\end{figure}

However, inspection of these bottom two coalescing bands indicates that they join only at the M point which is different from the other $\mathcal{PT}$ degeneracies whose overlapping
occurs for a range of wavevectors and depends on the non-Hermiticity factor $n_i$.  Therefore, this phenomenon is likely not directly related to the prediction of thresholdless $\mathcal{PT}$
transitions via the Heesh-Shubnikov theory.  Instead it results from the allowed symmetries at high-symmetry points determined by a \textit{reduction procedure}~\cite{sakoda2001}.  
The procedure begins by identifying the number of equivalent high-symmetry points in reciprocal space that are invariant under symmetry operations.  Fig.~\ref{rlmap} displays
a reciprocal lattice diagram with high symmetry points labeled. It is sufficient to work with the nominal unitary subgroup without the space group contribution, which here is 
just $C_{4v}$.  For example there is only one $\Gamma^{(1)}$ point (at the origin), and it is invariant under all symmetry operations in $C_{4v}$.  There are four $\Gamma^{(2)}$
points all of which are invariant under $E$, two of which are invariant under $\sigma_d$ and $\sigma_d'$ (each), and none of which are invariant under $C_2$, $C_4$, $C^{-1}_4$,
$\sigma_x$ and $\sigma_y$.  The point of interest in the present discussion is M$^{(2)}$, and there are eight equivalent points that are invariant under $E$.  M$^{(2)}$ is not
invariant under any other transformations in $C_{4v}$.  The ``character profile'' of M$^{(2)}$ will then be $P = (8,0,0,0,0,0,0,0)$ where the ordered set represents the instances of
the elements of $C_{4v}$ in the order $(E,C_2,C_4,C^{-1}_4,\sigma_x,\sigma_y,\sigma_d,\sigma_d')$.

One then calculates the projection of the possible irreducible representations onto the character profile of M$^{(2)}$.  This is done using the formula
\begin{equation}
\lambda_i = \frac{1}{h}\sum_{R \in C_{4v}} P(R) \chi_i(R) 
\end{equation}
where $h$ is the order of the group $C_{4v}$ ($h=8$), $P(R)$ represents the character profile of the element $R$ discussed above, $\chi_i(R)$ is the character of operation $R$ in the $i$th
irreducible representation of $C_{4v}$, and $\lambda_i$ counts the instances of the $i$th representation associated with M$^{(2)}$.  Performing this calculation for the 5 irreducible representations
of $C_{4v}$ results in one each of the four 1D representations and two instances of the lone 2D representation.  Because the 1D representations pair up to form
the 2D corepresntations in the complex-conjugate frequency regime, one expects two lone Type (c) corepresentations, and two 2D ``classically degenerate'' representations.  And this behavior is exhibited
in Fig.~\ref{kMzm2} where one sees the two Type (c) bands labeled $\Gamma_1$ and $\Gamma^{u}_3$.   The bottom two bands are also essentially Type (c) modes since they are in the complex-conjugate
regime, but due to the reduction procedure, they also exhibit behavior associated with ``classically'' doubly degenerate modes exactly at the M point.

This observation raises an interesting question: if the reduction procedure requires that two 2D classically degenerate modes appear at M$^{(2)}$, then shouldn't these
appear as pure 2D degenerate modes similar to modes (1,2) at M$^{(1)}$?  It appears that there are two competing effects at play: on the one hand, the reduction procedure provides
information based purely on the lattice (square in this case); whereas, the formation of gain and loss modes occurs within the lattice sites.  Apparently, the system favors
the formation of gain/loss modes (Type (c) corepresentations) even though the lattice symmetry demands that two 2D classically degenerate modes appear.  Ultimately, the phenomenon
at M$^{(2)}$ is the result of satisfying both constraints.

Based on observing the transformation properties of the fields in Fig.~\ref{kMzm2} we can write down the matrix representations of the operators associated with the degenerate set
of complex-conjugate modes.  The unitary operators $R$ transform according to 
\begin{equation}
R \begin{pmatrix}
H_1 \\
H_3 \\
H_2 \\
H_4 \\
\end{pmatrix} = 
\begin{pmatrix}
a_{11} & a_{12} & 0         & 0      \\
a_{21} & a_{22} & 0         & 0       \\
0         & 0          & a_{11} & a_{12} \\
0         & 0          & a_{21} & a_{22} \\
\end{pmatrix}
\begin{pmatrix}
H_1 \\
H_3 \\
H_2 \\
H_4 \\
\end{pmatrix}
\end{equation}

\noindent
where the fields are numbered according to the format in Fig.~\ref{kMzm2}.  This $4 \times 4$ matrix consists of two identical $2 \times 2$ matrices on the diagonal.
These $2 \times 2$ matrices are diagonal for $E, C_2, \sigma_d, \sigma_d'$ and antidiagonal for $C_4, C^{-1}_4, \sigma_x, \sigma_y$.
The antiunitary operators $R$ transform according to
\begin{equation}
R \begin{pmatrix}
H_1 \\
H_3 \\
H_2 \\
H_4 \\
\end{pmatrix} = 
\begin{pmatrix}
0         & 0          & a_{13} & a_{14} \\
0         & 0          & a_{23} & a_{24} \\
a_{13} & a_{14} & 0         & 0      \\
a_{23} & a_{24} & 0         & 0       \\
\end{pmatrix}
\begin{pmatrix}
H_1 \\
H_3 \\
H_2 \\
H_4 \\
\end{pmatrix}.
\end{equation}
Similarly, this $4 \times 4$ matrix consists of two identical $2 \times 2$ matrices but now on the antidiagonal.
These $2 \times 2$ matrices are diagonal for $\xi, \zeta, \mu_d, \mu_d'$ and antidiagonal for $\gamma, \gamma^{-1}, \mu_x, \mu_y$.
These operators correspond to the same elements of $C_{4v}$ listed above but combined with $\{\mathcal{T} | \frac{\Lambda}{2} \frac{\Lambda}{2} \}$ (see Ref.~\cite{smprx3} for more details).

 \section{IV. DISCUSSION AND  CONCLUSIONS}
 
 In this work we have shown how Heesh-Shubnikov group theory can be used to predict thresholdless $\mathcal{PT}$ transitions in 2D periodic photonic lattices that posses $\mathcal{PT}$ symmetry.
 While the theoretical framework was developed decades ago, we have found it instrumental in understanding the complex and varied modal behavior of these modern 
 materials.  Notably, we focussed on high symmetry points in the reciprocal lattice and found that at the $\Gamma$ point, one expects all modes to remain in the real-frequency
 regime; at the X point, one expects all modes to be in the complex-conjugate frequency regime; and at the M point, one expects modes in the complex-conjugate frequency
 regime and doubly degenerate modes in the real-frequency regime.  Interesting behavior was observed at the M point where modes possessed a combination of 2D degeneracy
 and complex-conjugate frequency formation.
 
 This work lays a theoretical foundation for important future studies.  Theoretically, a similar analysis applied to other lattice types as well as three-dimensional photonic crystals
 will likely uncover even more interesting behavior.  From an engineering perspective, the real frequency states at the $\Gamma$ point shows that these materials could be
 designed to create photonic bandgap devices such as waveguides and cavities.  The large density of states associated with the high-order degeneracy discovered at the M 
 point could be useful for bandedge lasers.  And the broad understanding of the band structure in general will prove critical for designing superprism-type devices and metasurfaces.

\bibliography{/Users/adammock1/Desktop/bib/myRef}
\end{document}


{\Large \textbf{\textit{Supplemental material for}} } \\

{\Large \textbf{Comprehensive understanding of tarity-time transitions in $\mathcal{PT}$ symmetric photonic crystals with an antiunitary group theory} } \\

\section{Symmetry elements and corepresentation tables for modes at $\Gamma$ point [$\mathbf{k} = 0$]} 

\subsection{Definition of symmetry operators}

The Heesh-Shubnikov little group at $\mathbf{k} = 0$ is $\mathcal{M}^{\mathbf{k} = 0} = C_{4v} + AC_{4v} = C_{4v} + \mathcal{T} \{ E | \frac{\Lambda}{2}\frac{\Lambda}{2}  \} C_{4v} $.
The elements of $C_{4v}$ are $C_{4v} = (E, C_2, C_4, C_4^{-1}, \sigma_x, \sigma_y, \sigma_d, \sigma_d')$.  $E$ is the identity.  $C_2$ is a rotation about the $z$ axis by $\pi$ radians.
$C_4$ ($C^{-1}_4$) is a rotation about the $z$ axis by $\pi/2$ ($-\pi/2$) radians.  $\sigma_x$ ($\sigma_y$) is mirror reflection through a plane passing through the origin and perpendicular to the
$x$ ($y$) axis.  $\sigma_d$ ($\sigma_d'$) is mirror reflection through a plane passing through the origin and parallel to $\boldsymbol{x}+\boldsymbol{y}$ ($-\boldsymbol{x}+\boldsymbol{y}$).
The antiunitary elements are 
$\mathcal{W} = \mathcal{T} \{ E | \frac{\Lambda}{2}\frac{\Lambda}{2}  \} C_{4v} = (\xi, \gamma_2, \gamma_4, \gamma_4^{-1}, \mu_x, \mu_y, \mu_d, \mu_d')$ where the greek
symbols correspond to the same ordered set shown for $C_{4v}$ but with each element multiplied by $\mathcal{T} \{ E | \frac{\Lambda}{2}\frac{\Lambda}{2} \}$. 

\subsection{Identification of corepresentations}

As discussed in the main text, at $\mathbf k=0$, representations of the space group do not need to be explicitly included since $\exp[i \mathbf{k} \cdot (m\boldsymbol{x}  + n\boldsymbol{y})\Lambda)] = 1$ for
all $m$ and $n$.   Performing the Dimmock-Wheeler test yields $\mathcal{W}^2 = ( E, E, E, E, E, E, C_2, C_2 )$ where the $C_2$ elements result from $(\gamma_4)^2 = (\gamma_4^{-1})^2=C_2$.
Table~\ref{charC4v1} shows the characters of the $C_{4v}$ irreducible representations and the result of the Dimmock-Wheeler test.  All representations are Type (a).  Therefore, no $\mathcal{PT}$ symmetry transitions
are expected to occur at $\mathbf{k}=0$, and all modes maintain real-valued eigenfrequencies.  


\begin{table}[ht]
\caption{Character table for $C_{4v}$ which is the unitary subgroup of $\mathcal{M}^{\mathbf{k} = 0}$. }
\begin{center}
\begin{ruledtabular}
\begin{tabular}{c|c c c c c | c c}
$C_{4v}$ & $E$ &$C_2$ & $C_4$, $C_4^{-1}$ & $\sigma_x$,$\sigma_y$ & $\sigma_d$,$\sigma_d'$ & $\alpha$ & Correp. Type \\
\hline
   $ A_1 $   & 1 & 1 & 1 & 1 & 1      &      8      &     (a)\\
   $ A_2 $   & 1 & 1 & 1 &-1 &-1      &      8      &     (a)\\
   $ B_1 $   & 1 & 1 &-1 & 1 &-1      &      8      &     (a)\\
   $ B_2 $   & 1 & 1 &-1 &-1 & 1      &      8      &     (a)\\
   $ E     $    & 2 &-2 & 0 & 0 & 0      &      8      &     (a)\\
\end{tabular}
\end{ruledtabular}
\end{center}
\label{charC4v1}
\end{table}

\subsection{Construction of corepresentations}

The components of the $i$th Type (a) corepresentation $\mathbf{\Gamma}_i$ for the unitary elements $R \in \mathcal{N}$ are given by 
$\mathbf{\Gamma}_i(R) = \mathbf{\Delta}_i(R)$ where $\mathbf{\Delta}_i(R)$ is the $i$th classical representation of $R$ in $\mathcal{N}$.
The components of the $i$th Type (a) corepresentation $\mathbf{\Gamma}_i$ for the antiunitary elements $R \in \mathcal{W}$ are given by
$\mathbf{\Gamma}_i(RA) = \boldsymbol{\beta} \mathbf{\Delta}_i(R)$ where $A \in \mathcal{W}$ is an arbitrary but fixed antiunitary operator and 
$\boldsymbol{\beta} \boldsymbol{\beta}^* = \mathbf{\Delta}_i(A^2)$~\cite{wigner1959, ElBatanouny2008}.  
In the determination of corepresentation for antiunitary elements, we will use $A = \xi $. The antiunitary corepresentations are calculated as follows.

$\mathbf{\Gamma}_i(E \xi)                 = \mathbf{\Gamma}_i(\xi)                           = \boldsymbol{\beta} \mathbf{\Delta}_i(E)              $ \\
$\mathbf{\Gamma}_i(C_2 \xi)            = \mathbf{\Gamma}_i(\gamma_2)              = \boldsymbol{\beta} \mathbf{\Delta}_i(C_2)          $ \\
$\mathbf{\Gamma}_i(C_4 \xi)            = \mathbf{\Gamma}_i(\gamma_4)              = \boldsymbol{\beta}\mathbf{\Delta}_i(C_4)           $ \\
$\mathbf{\Gamma}_i(C_4^{-1} \xi)    = \mathbf{\Gamma}_i(\gamma_4^{-1})       = \boldsymbol{\beta} \mathbf{\Delta}_i(C_4^{-1})   $ \\
$\mathbf{\Gamma}_i(\sigma_x \xi)   = \mathbf{\Gamma}_i(\mu_x)                     =  \boldsymbol{\beta} \mathbf{\Delta}_i(\sigma_x)   $ \\ 
$\mathbf{\Gamma}_i(\sigma_y \xi)    = \mathbf{\Gamma}_i(\mu_y)                    =  \boldsymbol{\beta} \mathbf{\Delta}_i(\sigma_y)   $ \\ 
$\mathbf{\Gamma}_i(\sigma_d \xi)   = \mathbf{\Gamma}_i(\mu_d)                    =  \boldsymbol{\beta} \mathbf{\Delta}_i(\sigma_d)   $ \\ 
$\mathbf{\Gamma}_i(\sigma_d' \xi)   = \mathbf{\Gamma}_i(\mu_d')                   =  \boldsymbol{\beta} \mathbf{\Delta}_i(\sigma_d')  $ \\

Using $A = \xi$ results in $\boldsymbol{\beta} \boldsymbol{\beta}^* = {\bf\Delta}_i(E)$.
For the one-dimensional corepresentations, $\beta \beta^* = \Delta_i(E) = 1$ (where boldface is removed to indicate 
scalar quantities) resulting in $\beta = \exp(\pm i \theta)$, and the total number of corepresentations is doubled.
Table~\ref{antiunitaryMk0} summarizes the one-dimensional corepresentations using $\beta = \pm 1$.  
For the 2D corepresentations, 

\[ \boldsymbol{\beta} \boldsymbol{\beta}^* = {\bf\Delta}_i(E) =  
\begin{pmatrix}
          1  & 0   \\
          0  & 1   \\
 \end{pmatrix}   \]

Matrices of the form

\[ \boldsymbol{\beta} = 
 \begin{pmatrix}
          e^{i \theta_1}  & 0                            \\
          0                          & e^{i\theta_2}   \\
 \end{pmatrix} 
\]

satisfy the equation for real $\theta_{1,2}$.

We note also that corepresentations related by unitary transformations ($U$) are equivalent.  However, the specific transformation mechanism is modified slightly to account for the anitunitarity of the operators.  Specifically,
the corepresentations $\boldsymbol{\Gamma}$ and $\boldsymbol{\Gamma}'$ are equivalent if they are related by $U \boldsymbol{\Gamma} U^{-1} = \boldsymbol{\Gamma}'$ for the unitary elements and 
$U \boldsymbol{\Gamma} (U^*)^{-1} = \boldsymbol{\Gamma}'$ for the antiunitary elements~\cite{wigner1959,ElBatanouny2008}.   

In the bandstructure shown in Fig. 2 of the main text, one band at the $\Gamma$ point is labeled $\Gamma_5^{\beta, u}$.  This notation indicates that the corepresentation at this point is given by
$\Gamma_5$ subject to unitary transformation by the matrix
\[ \boldsymbol{u} = 
 \begin{pmatrix}
          1  & 0    \\
          0  & -1  \\
 \end{pmatrix} 
\]
and the antiunitary corepresentation matrices are to be subsequently multiplied by 
\[ \boldsymbol{\beta} = 
 \begin{pmatrix}
         -1   &  0   \\
          0   & -1   \\
 \end{pmatrix}.
\]

\begin{table}[ht]
\caption{Corepresentation of unitary elements of $\mathcal{M}^{\vec k = 0}$. }
\begin{center}
\begin{ruledtabular}
\begin{tabular}{c|c c c c c}
$C_{4v}$ & $E$ &$C_2$ & $C_4$, $C_4^{-1}$ & $\sigma_x$,$\sigma_y$ & $\sigma_d$,$\sigma_d'$ \\
\hline
   $ A_1,  \mathbf{\Gamma}_1^{\pm} $   & 1 & 1 & 1 & 1 & 1      \\
   $ A_2,  \mathbf{\Gamma}_2^{\pm} $   & 1 & 1 & 1 &-1 &-1      \\
   $ B_1,  \mathbf{\Gamma}_3^{\pm} $   & 1 & 1 &-1 & 1 &-1      \\
   $ B_2,  \mathbf{\Gamma}_4^{\pm} $   & 1 & 1 &-1 &-1 & 1      \\
   $ E ,     \mathbf{\Gamma}_5^{\pm}  $   & 2 &-2 & 0 & 0 & 0      \\
   $ E,       \mathbf{\Gamma}_5     $    & 
   $\begin{pmatrix}
         1 & 0  \\
         0 & 1  \\
 \end{pmatrix} $ 
 &
 $\begin{pmatrix}
         -1 & 0  \\
         0 & -1  \\
 \end{pmatrix} $ 
 &
 $\begin{pmatrix}
         0 &  1  \\
        -1 &  0  \\
 \end{pmatrix} $, 
 $\begin{pmatrix}
         0 &  -1  \\
         1 &  0  \\
 \end{pmatrix} $
 &
 $\begin{pmatrix}
         0 & -1  \\
         -1 & 0  \\
 \end{pmatrix} $,
 $\begin{pmatrix}
         0 & 1  \\
        1 &  0  \\
 \end{pmatrix} $
 &
 $\begin{pmatrix}
          -1 &  0  \\
          0 &   1  \\
 \end{pmatrix} $,
 $\begin{pmatrix}
          1 &  0  \\
          0 & -1  \\
 \end{pmatrix} $  \\
\end{tabular}
\end{ruledtabular}
\end{center}
\label{unitaryMk0}
\end{table}

\begin{table}[ht]
\caption{Corepresentation of antiunitary elements of $\mathcal{M}^{\vec k = 0}$. }
\begin{center}
\begin{ruledtabular}
\begin{tabular}{c | c c c c c}
$C_{4v}$ & $\xi$ &$\gamma_2$ & $\gamma_4$, $\gamma_4^{-1}$ & $\mu_x$, $\mu_y$ & $\mu_d$, $\mu_d'$  \\
\hline
   $ A_1,  \mathbf{\Gamma}_1^{\pm} $   &  $\pm 1$ & $\pm 1$  & $\pm 1$  & $\pm 1$ & $\pm 1$      \\
   $ A_2,  \mathbf{\Gamma}_2^{\pm} $   &  $\pm 1$ & $\pm 1$  & $\pm 1$  & $\mp 1$ & $\mp 1$      \\
   $ B_1,  \mathbf{\Gamma}_3^{\pm} $   &  $\pm 1$ & $\pm 1$  & $\mp 1$  & $\pm 1$ & $\mp 1$      \\
   $ B_2,  \mathbf{\Gamma}_4^{\pm} $   &  $\pm 1$ & $\pm 1$  & $\mp 1$  & $\mp 1$ & $\pm 1$      \\
   $ E,       \mathbf{\Gamma}_5     $    & 
$\begin{pmatrix}
         1 & 0  \\
         0 & 1  \\
 \end{pmatrix} $ 
 &
 $\begin{pmatrix}
         -1 & 0  \\
         0 & -1  \\
 \end{pmatrix} $ 
 &
 $\begin{pmatrix}
         0 &  1  \\
        -1 &  0  \\
 \end{pmatrix} $, 
 $\begin{pmatrix}
         0 &  -1  \\
         1 &  0  \\
 \end{pmatrix} $
 &
 $\begin{pmatrix}
          0 & -1  \\
         -1 &  0  \\
 \end{pmatrix} $,
 $\begin{pmatrix}
         0 & 1  \\
         1 & 0  \\
 \end{pmatrix} $
 &
 $\begin{pmatrix}
         -1 &  0  \\
          0 &  1  \\
 \end{pmatrix} $,
 $\begin{pmatrix}
           1 & 0  \\
           0 & -1  \\
 \end{pmatrix} $  \\
\end{tabular}
\end{ruledtabular}
\end{center}
\label{antiunitaryMk0}
\end{table}

\newpage

%
%

\section{Symmetry elements and corepresentation tables for modes at X point ($\mathbf{k} = \frac{\pi}{\Lambda}\boldsymbol{x} $)} 

\subsection{Definition of symmetry operators}

As stated in the main text, the symmetry elements in the group of $\mathbf{k} = \frac{\pi}{\Lambda}\boldsymbol{x}$ include $(E, C_2, \sigma_x, \sigma_y)$ and
$ \mathcal{T} \{ E | \frac{\Lambda}{2},\frac{\Lambda}{2}  \} (E, C_2, \sigma_x, \sigma_y)$.  Because $\exp[i \mathbf{k} \cdot (\boldsymbol{x}m + \boldsymbol{y}n)\Lambda)] = \exp[i \pi m]$ 
which is 1 for $m$ even and $-1$ for $m$ odd, these two space group representations must be explicitly included in the group.  
We introduce an overbar to indicate $-1$.  Then the unitary subgroup elements are 

$e = \{E | 2m\Lambda, n\Lambda \}$ \\
$\overline{e} = \{E | 2m\Lambda+\Lambda, n\Lambda \}$ \\
$c = \{C_2 | 2m\Lambda, n\Lambda \}$ \\
$\overline{c} = \{C_2 | 2m\Lambda + \Lambda, n\Lambda\}$ \\
$m_x = \{\sigma_x | 2m\Lambda, n\Lambda \}$ \\
$\overline{m}_x = \{\sigma_x | 2m\Lambda + \Lambda, n\Lambda \}$ \\
$m_y = \{\sigma_y | 2m\Lambda, n\Lambda \}$ \\
$\overline{m}_y = \{\sigma_y | 2m\Lambda + \Lambda, n\Lambda \}$.  \\

The antiunitary group elements are 

$\xi = \{ \mathcal{T} E | 2m\Lambda+\frac{\Lambda}{2}, n\Lambda+\frac{\Lambda}{2} \} $ \\
$\overline{\xi} = \{ \mathcal{T} E | 2m\Lambda + \Lambda + \frac{\Lambda}{2}, n\Lambda+\frac{\Lambda}{2} \} $ \\
$\zeta = \{ \mathcal{T}C_2 | 2m\Lambda+\frac{\Lambda}{2}, n\Lambda + \frac{\Lambda}{2} \} $ \\
$\overline{\zeta} = \{ \mathcal{T}C_2 | 2m\Lambda + \Lambda + \frac{\Lambda}{2}, n\Lambda+\frac{\Lambda}{2} \} $ \\
$\mu_x = \{ \mathcal{T}\sigma_x | 2m\Lambda+\frac{\Lambda}{2}, n\Lambda +\frac{\Lambda}{2}\}$ \\
$\overline{\mu}_x = \{ \mathcal{T}\sigma_x | 2m\Lambda + \Lambda + \frac{\Lambda}{2}, n\Lambda + \frac{\Lambda}{2}\}$ \\
$\mu_y = \{  \mathcal{T} \sigma_y | 2m\Lambda+\frac{\Lambda}{2}, n\Lambda +\frac{\Lambda}{2}\}$ \\
$\overline{\mu}_y = \{ \mathcal{T} \sigma_y | 2m\Lambda + \Lambda + \frac{\Lambda}{2}, n\Lambda + \frac{\Lambda}{2}\}$.  \\

Inspection of the multiplication table shown in Table~\ref{d2h_mult} shows that the unitary subgroup $\mathcal{N}$ is isomorphic to the abelian group $D_{2h} (2mm)$.  Therefore, 
$\mathcal{M} = \mathcal{N} + A \mathcal{N} = \mathcal{N} + \mathcal{T} \{ E | \frac{\Lambda}{2}\frac{\Lambda}{2}  \} \mathcal{N}$ with $\mathcal{N} = D_{2h}$, $A = \xi$ and
$\mathcal{W} = \mathcal{T} \{ E | \frac{\Lambda}{2}\frac{\Lambda}{2}  \} D_{2h} = \mathcal{T} \{ E | \frac{\Lambda}{2}\frac{\Lambda}{2}  \}  
(e, \overline{e}, c, \overline{c}, m_x, \overline{m}_x, m_y, \overline{m}_y)$.
Performing the Dimmock-Wheeler test yields $\mathcal{W}^2 = ( \overline{e}, \overline{e}, e, e, e, e, \overline{e}, \overline{e} )$

\begin{table}[ht]
\caption{Multiplication table of unitary subgroup at $\mathbf{k} = \frac{\pi}{\Lambda}\boldsymbol{x}$ for 2D $\mathcal{PT}$ symmetric lattice.  It is isomorphic to the point group $D_{2h}$.  }
\begin{center}
\begin{tabular}{c | c | c | c | c | c | c | c | c |}

$D_{2h} (2mm)$ & \hspace{0.2cm} $ e $ \hspace{0.2cm} & \hspace{0.2cm} $ c $ \hspace{0.2cm}
                              & \hspace{0.2cm} $ m_y $ \hspace{0.2cm} & \hspace{0.2cm} $ m_x $ \hspace{0.2cm}
                              & \hspace{0.2cm} $ \overline{e} $ \hspace{0.2cm} & \hspace{0.2cm} $ \overline{c} $ \hspace{0.2cm}
                              & \hspace{0.2cm} $ \overline{m}_y $ \hspace{0.2cm} & \hspace{0.2cm} $ \overline{m}_x $ \hspace{0.2cm} \\
\hline
$ e $                         &  $ e $                        &  $ c $                           &  $ m_y $                  &  $ m_x  $                      &  $ \overline{e} $          &  $ \overline{c} $          &  $ \overline{m}_y $      &  $ \overline{m}_x $  \\
\hline
$ c  $                         &  $ c $                        &  $  e $                          &  $ m_x  $                 &  $  m_y $                     &  $\overline{c}  $          &  $ \overline{e} $           &  $ \overline{m}_x $     &  $ \overline{m}_y $  \\
\hline
$ m_y $                    &  $ m_y $                   & $ m_x  $                     &  $  e $                      &  $ c $                            & $ \overline{m}_y $      &  $ \overline{m}_x $     &  $ \overline{e}       $     &  $ \overline{e}        $ \\
\hline
$ m_x $                   &  $ m_x $                    &  $ m_y $                     &  $ c $                        &  $ e $                           &  $  \overline{m}_x $    &  $ \overline{m}_y $       &  $ \overline{c}       $    &  $ \overline{e}        $ \\
\hline
$\overline{e} $       &  $ \overline{e} $       &  $ \overline{c} $        &  $ \overline{m}_y $ &  $ \overline{m}_x $   &  $ e $                            &  $ c $                               &  $ m_y $                       &  $ m_x $                     \\
\hline
$ \overline{c} $       & $ \overline{c} $        &  $ \overline{e} $       &  $ \overline{m}_x $ &  $ \overline{m}_y $   &  $ c $                             &  $ e $                              &  $ m_x $                       &  $ m_y $   \\
\hline
$ \overline{m}_y $ &  $ \overline{m}_y $  & $ \overline{m}_x $   & $ \overline{e} $      &  $ \overline{c} $         &  $ m_y  $                       &  $ m_x $                        &  $  e $                            &   $ c $ \\
\hline
$ \overline{m}_x $ & $ \overline{m}_x $   & $ \overline{m}_y $   &  $ \overline{c} $      & $ \overline{e} $         &  $ m_x $                        &  $ m_y  $                        &  $ c  $                           &   $  e $ \\
\hline
\end{tabular}
\end{center}
\label{d2h_mult}
\end{table}



%

\subsection{Character table of unitary subgroup}

Table~\ref{d2h_char} shows the character table for the $D_{2h}$ point group.  Only the bottom four irreducible representations are physically applicable in the present
scenario since the character must change sign between an operator and its overbarred counterpart. 

\begin{table}[ht]
\caption{Character table of the unitary subgroup of $\mathbf{k} = \frac{\pi}{\Lambda}\boldsymbol{x}$ which is isomorphic to $D_{2h} (mmm)$.}
\begin{center}
\begin{ruledtabular}
\begin{tabular}{c|c c c c c c c c | c c}
$D_{2h}$ & $ e $ & $ c $ & $ m_y $ & $ m_x $ & $ \overline{e} $ & $ \overline{c}$ & $ \overline{m}_y $ & $ \overline{m}_x $ & $\alpha$ & Correp. Type \\
\hline
   $ A_g      $   & 1 &  1 &  1  &  1 &   1 &  1 &  1 &  1  &      8      &     (a)\\
   $ B_{1g} $   & 1 &  1 & -1  & -1 &  1 &  1 & -1 & -1  &      8      &     (a)\\
   $ B_{2g} $   & 1 & -1 &  1  & -1 &  1 & -1 &  1 & -1  &      8      &     (a)\\
   $ B_{3g} $   & 1 & -1 & -1  &  1 &  1 & -1 & -1 &  1  &      8      &     (a)\\
   $ A_u      $   & 1 &  1 &  1   &  1 & -1 & -1 &-1 & -1  &      0      &     (c)\\
   $ B_{1u} $   & 1 &  1 & -1  & -1 & -1 & -1 &  1 &  1  &      0      &     (c)\\
   $ B_{2u} $   & 1 & -1 &  1  & -1 & -1 &  1 & -1 &  1  &      0      &     (c)\\
   $ B_{3u} $   & 1 & -1 & -1  &  1 & -1 &  1 &  1 & -1  &      0      &     (c)\\
\end{tabular}
\end{ruledtabular}
\end{center}
\label{d2h_char}
\end{table}

\subsection{Construction of corepresentations}

As stated in the main text, the components of the $i$th Type (c) corepresentation $\boldsymbol{\Gamma}_i$ for the unitary elements 
 $R \in \mathcal{N}$ are given by~\cite{wigner1959, ElBatanouny2008}

\begin{equation}
\boldsymbol{\Gamma}_i(R) = \begin{pmatrix}
    \boldsymbol{\Delta}(R)  &  \mathbf{0} \\
    \mathbf{0}               &   \boldsymbol{\Delta}^*(S^{-1}RS) 
\end{pmatrix}
\label{eqcorepU}
\end{equation}

\noindent
where $A = S\mathcal{T}$.   In the following we use $A = \xi = S\mathcal{T}$.  
Therefore, $S = \{E|2m \Lambda + \Lambda/2, n\Lambda + \Lambda/2 \}$ and $S^{-1} = \{E|-2m\Lambda - \Lambda/2, -n\Lambda - \Lambda/2 \}$.
Calculations needed for Eq.~\ref{eqcorepU} are provided in Table~\ref{eleUtc}.

\begin{table}[ht]
\caption{Determination of Type (c) corepresentations for unitary elements. }
\begin{center}
\begin{tabular}{c | c | c}
$R$                          &      $RS$  &  $S^{-1}RS$ \\
\hline
$e$                          &      $ \{E|2m\Lambda + \Lambda/2, n\Lambda + \Lambda/2 \}$                               & $\{E|0,0 \} = e$   \\
\hline
$\overline{e}$          &      $ \{E|2m\Lambda - \Lambda/2, n\Lambda - \Lambda/2 \}$                                   & $\{E|-\Lambda,-\Lambda \} = \overline{e}$   \\
\hline
$c$                          &      $ \{C_2 | 2m\Lambda + \Lambda + \Lambda/2, n\Lambda + \Lambda/2 \}$       & $\{C_2 | \Lambda, 0 \} = \overline{c}$   \\
\hline
$\overline{c}$          &      $ \{C_2 | 2m\Lambda + \Lambda/2, n\Lambda + \Lambda/2 \} $                          & $\{C_2 | 0, 0 \} = c$   \\
\hline
$m_x $                    &      $ \{\sigma_x | 2m\Lambda - \Lambda/2, n\Lambda + \Lambda/2 \}$                  &  $\{\sigma_x | -\Lambda, 0 \} = \overline{m}_x$\\
\hline
$\overline{m}_x $    &      $ \{\sigma_x | 2m\Lambda + \Lambda/2, n\Lambda + \Lambda/2 \}$                   &  $\{\sigma_x | 0, 0  \} = m_x$ \\
\hline
$m_y $                    &      $ \{\sigma_y | 2m\Lambda + \Lambda/2, n\Lambda - \Lambda/2 \}$                  &  $\{\sigma_y | 0, -\Lambda \} = m_y$\\
\hline
$\overline{m}_y $   &      $ \{\sigma_y | 2m\Lambda + \Lambda + \Lambda/2, n\Lambda - \Lambda/2 \}$   &  $\{\sigma_y | \Lambda, -\Lambda  \} = \overline{m}_y$ \\
\end{tabular}
\end{center}
\label{eleUtc}
\end{table}

The components of the $i$th Type (c) correpresentation $\boldsymbol{\Gamma}_i$ for the antiunitary elements $R \in \mathcal{W}$ are given by

\begin{equation}
\boldsymbol{\Gamma}_i(R) = \begin{pmatrix}
   \mathbf{0}                    &  \boldsymbol{\Delta}^*(A^{-1}R) \\
   \boldsymbol{\Delta}(RA)   &  \mathbf{0}  
\end{pmatrix}.
\label{eqcorepAU}
\end{equation}

With $A = \xi$ and $A^{-1} = \overline{\xi}$, the resulting calculations are provided in Table~\ref{eleAUtc}. \\

\begin{table}[ht]
\caption{Determination of Type (c) corepresentations for antiunitary elements.  }
\begin{center}
\begin{tabular}{c | c | c }
$B$                              &      $BA                            $  & $ A^{-1}B              $  \\
\hline
$\xi$                            &      $ \overline{e}           $  & $ e                         $   \\
\hline
$\zeta$                        &      $ c                             $  & $ c                         $   \\
\hline
$\mu_y $                    &      $  \overline{m}_y   $   &  $ m_y                       $   \\
\hline
$\mu_x $                    &      $  m_x                      $   &  $ m_x   $   \\
\hline
$\overline{\xi}$          &      $ e                             $   & $ \overline{e}     $   \\
\hline
$\overline{\zeta} $     &      $ \overline{c}           $   & $  \overline{c}     $   \\
\hline
$\overline{\mu}_y $  &      $  m_y                     $   &  $ \overline{m}_y     $ \\
\hline
$\overline{\mu}_x $  &      $  \overline{m}_x   $   &  $ \overline{m}_x     $ \\
\end{tabular}
\end{center}
\label{eleAUtc}
\end{table}

\begin{table}[ht]
\caption{ Corepresentations of the group of $k=\hat{x}\frac{\pi}{a}$ for the unitary symmetry operators.   }
\begin{center}
\begin{ruledtabular}
\begin{tabular}{c c | c c c c c c c  c }
Type  &  $D_{2h} (mmm)$ & $ e $ & $ c $ & $ m_y $ & $ m_x $ & $ \overline{e} $ & $ \overline{c} $ & $ \overline{m}_y $ & $ \overline{m}_x $ \\
\hline
  (a)  &  $ A_g,       \mathbf{\Gamma}_1 $       & 1 &  1 &  1  &  1 &  1 &   1  &   1 &  1      \\
  (a)  &  $ B_{1g},  \mathbf{\Gamma}_2 $       & 1 &  1 & -1  & -1 &  1 &   1  & -1 & -1     \\
  (a)  &  $ B_{2g},  \mathbf{\Gamma}_3 $       & 1 & -1 &  1  & -1 &  1 &  -1  &  1 & -1     \\
  (a)  &  $ B_{3g},  \mathbf{\Gamma}_4 $       & 1 &  -1 & -1 &  1 &  1 &  -1  & -1 &  1    \\
  (c)  &  $ A_u,        \mathbf{\Gamma}_5 $       &  
  
$\left( \begin{tabular}{cc}
         1 & 0  \\
         0 & 1  \\
        \end{tabular}
        \right)$
&
$\left( \begin{tabular}{cc}
         1 & 0  \\
         0 & -1  \\
        \end{tabular}
        \right)$
& 
$\left( \begin{tabular}{cc}
         1 & 0  \\
         0 & 1  \\
        \end{tabular}
        \right)$
&
$\left( \begin{tabular}{cc}
         1 & 0  \\
         0 & -1  \\
        \end{tabular}
        \right)$
&
$\left( \begin{tabular}{cc}
         -1 & 0  \\
         0 &  -1  \\
        \end{tabular}
        \right)$
&
$\left( \begin{tabular}{cc}
         -1 & 0  \\
         0 & 1  \\
        \end{tabular}
        \right)$
& 
$\left( \begin{tabular}{cc}
         -1 & 0  \\
         0 & -1  \\
        \end{tabular}
        \right)$
&
$\left( \begin{tabular}{cc}
         -1 & 0  \\
         0 &  1  \\
        \end{tabular}
        \right)$
\\

  (c)  &  $ B_{1u},  \mathbf{\Gamma}_6 $       & 
  
$\left( \begin{tabular}{cc}
         1 & 0  \\
         0 & 1  \\
        \end{tabular}
        \right)$
&
$\left( \begin{tabular}{cc}
          1 & 0  \\
         0 & -1  \\
        \end{tabular}
        \right)$
& 
$\left( \begin{tabular}{cc}
         -1 & 0  \\
         0 & -1  \\
        \end{tabular}
        \right)$
&
$\left( \begin{tabular}{cc}
         -1 & 0  \\
         0 & 1  \\
        \end{tabular}
        \right)$
&
$\left( \begin{tabular}{cc}
         -1 & 0  \\
         0 & -1  \\
        \end{tabular}
        \right)$
&
$\left( \begin{tabular}{cc}
         -1 & 0  \\
         0 & 1  \\
        \end{tabular}
        \right)$
& 
$\left( \begin{tabular}{cc}
          1 & 0  \\
         0 & 1  \\
        \end{tabular}
        \right)$
&
$\left( \begin{tabular}{cc}
         1 & 0  \\
         0 & -1  \\
        \end{tabular}
        \right)$
\\  

  (c)  &  $ B_{2u},  \mathbf{\Gamma}_7 $       & 
  
  $\left( \begin{tabular}{cc}
         1 & 0  \\
         0 & 1  \\
        \end{tabular}
        \right)$
&
$\left( \begin{tabular}{cc}
         -1 & 0  \\
         0 &  1  \\
        \end{tabular}
        \right)$
& 
$\left( \begin{tabular}{cc}
         1 & 0  \\
         0 & 1  \\
        \end{tabular}
        \right)$
&
$\left( \begin{tabular}{cc}
         -1 & 0  \\
         0 &  1  \\
        \end{tabular}
        \right)$
&
$\left( \begin{tabular}{cc}
         -1 & 0  \\
          0 & -1  \\
        \end{tabular}
        \right)$
&
$\left( \begin{tabular}{cc}
         1 &  0  \\
         0 & -1  \\
        \end{tabular}
        \right)$
& 
$\left( \begin{tabular}{cc}
          -1 & 0  \\
          0 & -1  \\
        \end{tabular}
        \right)$
&
$\left( \begin{tabular}{cc}
         1 & 0  \\
         0 & -1  \\
        \end{tabular}
        \right)$
\\

  (c)  &  $ B_{3u},  \mathbf{\Gamma}_8 $       & 

$\left( \begin{tabular}{cc}
         1 & 0  \\
         0 & 1  \\
        \end{tabular}
        \right)$
&
$\left( \begin{tabular}{cc}
         -1 & 0  \\
         0 & 1  \\
        \end{tabular}
        \right)$
& 
$\left( \begin{tabular}{cc}
         -1 & 0  \\
         0 & -1  \\
        \end{tabular}
        \right)$
&
$\left( \begin{tabular}{cc}
         1 & 0  \\
         0 & -1  \\
        \end{tabular}
        \right)$
&
$\left( \begin{tabular}{cc}
         -1 & 0  \\
         0 & -1  \\
        \end{tabular}
        \right)$
&
$\left( \begin{tabular}{cc}
         1 & 0  \\
         0 & -1  \\
        \end{tabular}
        \right)$
& 
$\left( \begin{tabular}{cc}
         1 & 0  \\
         0 & 1  \\
        \end{tabular}
        \right)$
&
$\left( \begin{tabular}{cc}
         -1 & 0  \\
         0 & 1  \\
        \end{tabular}
        \right)$
\\
\end{tabular}
\end{ruledtabular}
\end{center}
\label{unitaryMkX}
\end{table}

\begin{table}[ht]
\caption{ Corepresentations of the group of $k=\hat{x}\frac{\pi}{a}$ for the anti-unitary symmetry operators.   }
\begin{center}
\begin{ruledtabular}
\begin{tabular}{c c | c c c c c c c  c }
Type  &  $D_{2h} (mmm)$ & $\xi$                       & $ \zeta$                      & $ \mu_y $                    & $ \mu_x$ 
                                                & $ \overline{\xi} $  & $ \overline{\zeta} $   & $ \overline{\mu}_y $  & $ \overline{\mu}_x $ \\
\hline
  (a)  &  $ A_g,       \mathbf{\Gamma}_1 $       & 1 &  1 &  1  &  1 &  1 &   1  &   1 &  1      \\
  (a)  &  $ B_{1g},  \mathbf{\Gamma}_2 $       & 1 &  1 & -1  & -1 &  1 &   1  & -1 & -1     \\
  (a)  &  $ B_{2g},  \mathbf{\Gamma}_3 $       & 1 & -1 &  1  & -1 &  1 &  -1  &  1 & -1     \\
  (a)  &  $ B_{3g},  \mathbf{\Gamma}_4 $       & 1 &  -1 & -1 &  1 &  1 &  -1  & -1 &  1    \\
  (c)  &  $ A_u,       \mathbf{\Gamma}_5 $       &  
  
$\left( \begin{tabular}{cc}
         0 & 1  \\
         -1 & 0  \\
        \end{tabular}
        \right)$
&
$\left( \begin{tabular}{cc}
         0 &  1  \\
         1 & 0  \\
        \end{tabular}
        \right)$
& 
$\left( \begin{tabular}{cc}
         0 & 1  \\
         -1 & 0  \\
        \end{tabular}
        \right)$
&
$\left( \begin{tabular}{cc}
         0 & 1  \\
         1 & 0  \\
        \end{tabular}
        \right)$
&
$\left( \begin{tabular}{cc}
         0 & -1  \\
         1 &  0  \\
        \end{tabular}
        \right)$
&
$\left( \begin{tabular}{cc}
         0 & -1  \\
         -1 & 0  \\
        \end{tabular}
        \right)$
& 
$\left( \begin{tabular}{cc}
         0 & -1  \\
         1 & 0  \\
        \end{tabular}
        \right)$
&
$\left( \begin{tabular}{cc}
           0 & -1  \\
         -1 &  0  \\
        \end{tabular}
        \right)$
\\

  (c)  &  $ B_{1u},  \mathbf{\Gamma}_6 $       & 
  
$\left( \begin{tabular}{cc}
          0 & 1  \\
        -1 & 0  \\
        \end{tabular}
        \right)$
&
$\left( \begin{tabular}{cc}
         0 & 1  \\
         1 & 0  \\
        \end{tabular}
        \right)$
& 
$\left( \begin{tabular}{cc}
          0 & -1  \\
         1 &  0 \\
        \end{tabular}
        \right)$
&
$\left( \begin{tabular}{cc}
          0 & -1  \\
         -1 & 0  \\
        \end{tabular}
        \right)$
&
$\left( \begin{tabular}{cc}
         0 & -1  \\
         1 & 0  \\
        \end{tabular}
        \right)$
&
$\left( \begin{tabular}{cc}
         0 & -1  \\
        -1 & 0  \\
        \end{tabular}
        \right)$
& 
$\left( \begin{tabular}{cc}
          0 & 1  \\
         -1 & 0  \\
        \end{tabular}
        \right)$
&
$\left( \begin{tabular}{cc}
         0 & 1  \\
         1 & 0  \\
        \end{tabular}
        \right)$
\\  

  (c)  &  $ B_{2u},  \mathbf{\Gamma}_7 $       & 
  
  $\left( \begin{tabular}{cc}
         0 & 1  \\
         -1 & 0  \\
        \end{tabular}
        \right)$
&
$\left( \begin{tabular}{cc}
         0 & -1  \\
         -1 &  0  \\
        \end{tabular}
        \right)$
& 
$\left( \begin{tabular}{cc}
         0 & 1  \\
         -1 & 0  \\
        \end{tabular}
        \right)$
&
$\left( \begin{tabular}{cc}
         0 & -1  \\
         -1 & 0  \\
        \end{tabular}
        \right)$
&
$\left( \begin{tabular}{cc}
          0 & -1  \\
          1 & 0  \\
        \end{tabular}
        \right)$
&
$\left( \begin{tabular}{cc}
         0 &  1  \\
         1 &  0 \\
        \end{tabular}
        \right)$
& 
$\left( \begin{tabular}{cc}
          0 & -1  \\
          1 & 0  \\
        \end{tabular}
        \right)$
&
$\left( \begin{tabular}{cc}
         0 &  1  \\
         1 & 0  \\
        \end{tabular}
        \right)$
\\

  (c)  &  $ B_{3u},  \mathbf{\Gamma}_8 $       & 

$\left( \begin{tabular}{cc}
         0 & 1  \\
        -1 & 0  \\
        \end{tabular}
        \right)$
&
$\left( \begin{tabular}{cc}
          0 & -1  \\
         -1 & 0  \\
        \end{tabular}
        \right)$
& 
$\left( \begin{tabular}{cc}
         0 & -1  \\
         1 & 0  \\
        \end{tabular}
        \right)$
&
$\left( \begin{tabular}{cc}
         0 & 1  \\
         1 & 0  \\
        \end{tabular}
        \right)$
&
$\left( \begin{tabular}{cc}
         0 & -1  \\
         1 & 0  \\
        \end{tabular}
        \right)$
&
$\left( \begin{tabular}{cc}
         0 & 1  \\
         1 & 0  \\
        \end{tabular}
        \right)$
& 
$\left( \begin{tabular}{cc}
         0 & 1  \\
         -1 & 0  \\
        \end{tabular}
        \right)$
&
$\left( \begin{tabular}{cc}
         0 & -1  \\
        -1 & 0  \\
        \end{tabular}
        \right)$
\\
  
\end{tabular}
\end{ruledtabular}
\end{center}
\label{antiunitaryMkX}
\end{table}

\subsection{Corepresentations equivalent through unitary transformation}

Corepresentations $\mathbf{\Gamma}_5$ and $\mathbf{\Gamma}_7$ and corepresentations $\mathbf{\Gamma}_6$ and $\mathbf{\Gamma}_8$ are equivalent through a similarity transformation using the unitary matrix:
\begin{displaymath}
U = \left( \begin{array}{cc}
                             0 & -e^{-i \theta} \\
                             e^{i \theta} & 0 \\
\end{array} \right)
\end{displaymath}
for real $\theta$.  When the two-dimensional corepresentations are constructed from nominally one-dimensional representations, the total number of dimensions will be preserved.
Therefore, even though we get four corepresentations from four one-dimensional representations, only two of them will be unique since two two-dimensional corepresentaitons yield
four total dimensions consistent with the preliminary four one-dimensional representations.  Therefore, the presence of only two unique corepresentations among $\mathbf{\Gamma}_5$ - $\mathbf{\Gamma}_8$
is expected.

In the bandstructure shown in Fig. 2 of the main text, the corepresentation labels at the X point include superscripts $u_1$ and $u_2$.  
This notation indicates that the corepresentation matrices given in Tables~\ref{unitaryMkX} and \ref{antiunitaryMkX} should be transformed using 
\[ u_1 = 
 \begin{pmatrix}
          c  & 0    \\
          0  & -ic^*  \\
 \end{pmatrix} 
\]
\[ u_2 = 
 \begin{pmatrix}
          c  & 0    \\
          0  & ic^*  \\
 \end{pmatrix} 
\]
where $c$ is a complex constant of unity modulus.

\newpage

\section{ Symmetry operators and corepresentation tables at the M point [$\mathbf{k} = (\boldsymbol{x} + \boldsymbol{y}) \frac{\pi}{\Lambda} $] } 

\subsection{Definition of symmetry operators}

Now consider $\mathbf{k} = (\boldsymbol{x} + \boldsymbol{y}) \frac{\pi}{\Lambda} $ which is the M point.  The symmetry elements in the group of 
$\mathbf{k}$ include $(E, C_2, C_4, C_4^{-1}, \sigma_x, \sigma_y, \sigma_d, \sigma_d')$ and
$ \mathcal{T} \{ E | \frac{\Lambda}{2},\frac{\Lambda}{2}  \} (E, C_2, C_4, C_4^{-1}, \sigma_x, \sigma_y, \sigma_d, \sigma_d')$.  
Because $\exp[i \mathbf{k} \cdot (\boldsymbol{x}m + \boldsymbol{y}n)\Lambda] = \exp[i \pi (m+n)]$ which is 1 for $m+n$ even and
$-1$ for $m+n$ odd, these two space group representations must be explicitly included in the group.  To represent the even 
combination, we write $e = \{E | 2m\Lambda, 2n\Lambda \}$. The odd combination can arise in two ways:
$\{E | (2m+1)\Lambda, n\Lambda \}$ or $\{E | 2m\Lambda, (2n+1)\Lambda \}$.  Both forms produce the same results in the analysis, so we will simply use the overbar symbol to denote $m+n$ odd: $\overline{e}$.
Similarly

$e = \{E | 2m\Lambda, 2n\Lambda \}$ \\
$C_2 = \{C_2 | 2m\Lambda, 2n\Lambda \}$ \\
$C_4 = \{C_4 | 2m\Lambda, 2n\Lambda \}$ \\
$C_4^{-1} = \{C_4^{-1} | 2m\Lambda, 2n\Lambda \}$ \\
$m_x = \{\sigma_x | 2m\Lambda, 2n\Lambda \}$ \\
$m_y = \{\sigma_y | 2m\Lambda, 2n\Lambda \}$ \\
$\sigma_d = \{\sigma_d | 2m\Lambda, 2n\Lambda \}$ \\
$\sigma_d' = \{\sigma_d' | 2m\Lambda, 2n\Lambda \}$ 

with corresponding $\overline{e}$, $\overline{C}_2$, $\overline{C}_4$, $\overline{C}_4^{-1}$, $\overline{m}_x$, $\overline{m}_y$, $\overline{\sigma}_d$, $\overline{\sigma}_d'$.

The anti-unitary group elements are 

$\xi = \{ \mathcal{T} E | 2m\Lambda+\frac{\Lambda}{2}, 2n\Lambda+\frac{\Lambda}{2} \} $ \\
$\zeta = \{ \mathcal{T}C_2 | 2m\Lambda+\frac{\Lambda}{2}, 2n\Lambda + \frac{\Lambda}{2} \} $ \\
$\gamma = \{ \mathcal{T}C_4 | 2m\Lambda+\frac{\Lambda}{2}, 2n\Lambda + \frac{\Lambda}{2} \} $ \\
$\gamma^{-1} = \{ \mathcal{T}C_4^{-1} | 2m\Lambda+\frac{\Lambda}{2}, 2n\Lambda + \frac{\Lambda}{2} \} $ \\
$\mu_x = \{ \mathcal{T}\sigma_x | 2m\Lambda+\frac{\Lambda}{2}, 2n\Lambda +\frac{\Lambda}{2}\}$ \\
$\mu_y = \{  \mathcal{T} \sigma_y | 2m\Lambda+\frac{\Lambda}{2}, 2n\Lambda +\frac{\Lambda}{2}\}$ \\
$\mu_d = \{ \mathcal{T}\sigma_d | 2m\Lambda+\frac{\Lambda}{2}, 2n\Lambda +\frac{\Lambda}{2}\}$ \\
$\mu_d' = \{  \mathcal{T} \sigma_d' | 2m\Lambda+\frac{\Lambda}{2}, 2n\Lambda +\frac{\Lambda}{2}\}$  

with corresponding $\overline{\xi}$, $\overline{\zeta}$, $\overline{\gamma}$, $\overline{\gamma}^{-1}$, $\overline{\mu}_x$, $\overline{\mu}_y$, $\overline{\mu}_d$, $\overline{\mu}_d'$.

Inspection of the multiplication table displayed in Table~\ref{d4h_mult} shows that the unitary subgroup is isomorphic to the group $D_{4h}  (4/mmm)$.  
The HSLG at the M point is then $\mathcal{M}^{\mathbf{k} = (\boldsymbol{x} + \boldsymbol{y}) \frac{\pi}{\Lambda}} = D_{4h} + \{ \mathcal{T} | \frac{\Lambda}{2}\frac{\Lambda}{2} \} D_{4h} $. 
The set of antiunitary operators is given by $\mathcal{W} = \{\mathcal{T}E | \frac{\Lambda}{2}\frac{\Lambda}{2}  \} D_{4h} $.

Using the ordering $\mathcal{W} = (\xi, \overline{\xi}, \gamma, \overline{\gamma}, \gamma^{-1}, \overline{\gamma}_4^{-1}, \zeta, \overline{\zeta}, \mu_x, 
\overline{\mu}_x, \mu_y, \overline{\mu}_y, \mu_d, \overline{\mu}_d, \mu_d',  \overline{\mu}_d')$
the Dimmock-Wheeler test yields $\mathcal{W}^2 = ( e, e, \overline{C}_2, \overline{C_2}, \overline{C_2}, \overline{C_2}, e, e, \overline{e}, \overline{e}, \overline{e}, \overline{e}, e, e, e, e)$.


\begin{sidewaystable}[ht]
\caption{Multiplication table of unitary subgroup at $\mathbf{k}= (\boldsymbol{x} + \boldsymbol{y})\frac{\pi}{\Lambda} $ for 2D $\mathcal{PT}$ symmetric lattice.  It is isomorphic to the point group $D_{4h}$.  }
\begin{center}
\begin{tabular}{c | c | c | c | c | c | c | c | c | c | c | c | c | c | c | c | c |}

$D_{4h} (4/mmm)$ & \hspace{0.2cm} $ e      $ \hspace{0.2cm}               & \hspace{0.2cm} $ \overline{e} $ \hspace{0.2cm}
                              & \hspace{0.2cm} $  C_4 $ \hspace{0.2cm}               & \hspace{0.2cm} $  \overline{C}_4 $ \hspace{0.2cm}
                              & \hspace{0.2cm} $ C_4^{-1} $ \hspace{0.2cm}        & \hspace{0.2cm} $ \overline{C}_4^{-1} $ \hspace{0.2cm}
                              & \hspace{0.2cm} $ C_2   $ \hspace{0.2cm}              & \hspace{0.2cm} $ \overline{C}_2 $ \hspace{0.2cm} 
                              & \hspace{0.2cm} $ m_x   $ \hspace{0.2cm}              & \hspace{0.2cm} $ \overline{m}_x $ \hspace{0.2cm}
                              & \hspace{0.2cm} $ m_y    $ \hspace{0.2cm}             & \hspace{0.2cm} $  \overline{m}_y $ \hspace{0.2cm}
                              & \hspace{0.2cm} $ \sigma_d $ \hspace{0.2cm}       & \hspace{0.2cm} $ \overline{\sigma}_d $ \hspace{0.2cm}
                              & \hspace{0.2cm} $ \sigma_d'     $ \hspace{0.2cm}   & \hspace{0.2cm} $ \overline{\sigma}_d' $ \hspace{0.2cm} \\
\hline
$ e $                                      &  $ e $                                  &  $ \overline{e} $                 &  $ C_4 $                             &  $ \overline{C}_4  $             &  $ C_4^{-1}  $                  &  $ \overline{C}_4^{-1} $  &  $ C_2 $                             &  $ \overline{C}_2 $            &  $ m_x $                               &  $\overline{m}_x$            &  $ m_y $                         &  $\overline{m}_y $          & $ \sigma_d $                     & $ \overline{\sigma}_d $  &  $ \sigma_d' $                     & $ \overline{\sigma}_d' $   \\
\hline
$ \overline{e} $                   &  $ \overline{e}  $               &  $  e $                                  &  $ \overline{C}_4  $          &  $  C_4 $                               &  $\overline{C}_4^{-1}  $ &  $ C_4^{-1}  $                    &  $  \overline{C}_2 $         &  $ C_2 $                              & $ \overline{m}_x $             &  $ m_x $                             &  $ \overline{m}_y $       &  $ m_y $                           &  $ \overline{\sigma}_d $  & $ \sigma_d $                     &  $ \overline{\sigma}_d' $  & $ \sigma_d' $   \\      
\hline
$ C_4 $                                &  $ C_4 $                             & $ \overline{C}_4  $           &  $  C_2  $                            &  $ \overline{C}_2  $            & $ e$                                    &  $ \overline{e} $               &  $ C_4^{-1} $                     &  $ \overline{C}_4^{-1}  $  & $ \sigma_d' $                     &  $ \overline{\sigma}_d' $ & $\sigma_d $                   & $\overline{\sigma}_d $  & $m_x $                                &  $\overline{m}_x $          &  $ m_y $                              & $\overline{m}_y $ \\
\hline
$  \overline{C}_4 $             &  $\overline{C}_4 $            &  $ m_y $                             &  $ \overline{C}_2 $            &  $ C_2 $                               &  $  \overline{e} $              &  $ e $                                  &  $ \overline{C}_4^{-1} $  &  $ C_4^{-1} $                      & $ \overline{\sigma}_d' $  &  $ \sigma_d' $                   &$\overline{\sigma}_d$   & $\sigma_d $                    & $ \overline{m}_x $             & $ m_x $                             & $ \overline{m}_y $            & $m_y$ \\
\hline
$C_4^{-1}  $                        &  $ C_4^{-1}  $                    &  $ \overline{C}_4^{-1} $  &  $ e $                                    &  $ \overline{e} $                  &  $ C_2 $                            &  $ \overline{C}_2 $          &  $ C_4 $                             &  $ \overline{C}_4 $            & $ \sigma_d $                     & $ \overline{\sigma_d} $  & $\sigma_d' $                   & $\overline{\sigma}_d'$  & $m_y$                                 & $ \overline{m}_y $           & $ m_x $                              & $ \overline{m}_x$ \\
\hline 
$ \overline{C}_4^{-1} $     & $  \overline{C}_4^{-1} $   &  $ C_4 $                             &  $ \overline{e} $                  &  $ e $                                    &  $\overline{C}_2 $           &  $ C_2 $                            &  $ \overline{C}_4 $           &  $ C_4 $                              & $ \overline{\sigma}_d $   & $ \sigma_d $                    & $\overline{\sigma}_d' $ & $ \sigma_d' $                  & $\overline{m}_y $             & $ m_y $                              & $ \overline{m}_x $            & $m_x$ \\
\hline
$C_2 $                                 &  $ C_2 $                             & $ \overline{C}_2 $            & $ C_4^{-1} $                        &  $ \overline{C}_4^{-1} $   &  $C_4  $                            &  $ \overline{C}_4^{-1} $  &  $  e $                                 &   $ \overline{e} $                & $m_y $                               & $\overline{m}_y$              & $m_x $                             & $\overline{m}_x $          & $\sigma_d'$                       & $\overline{\sigma}_d' $  &  $ \sigma_d $                     & $\overline{\sigma}_d $ \\
\hline
$ \overline{C}_2 $              & $ \overline{C}_2 $            & $ C_2 $                             &  $ \overline{C}_4^{-1} $     & $ C_4^{-1} $                      &  $ \overline{C}_4 $          &  $ C_4  $                            &  $ \overline{e}  $               &   $  e $                                 & $ \overline{m}_y $           &  $m_y$                                & $\overline{m}_x $          & $m_x $                             & $\overline{\sigma}_d' $    & $\sigma_d'$                     & $ \overline{\sigma}_d $    & $\sigma_d$ \\
\hline
$ m_x $                                &  $ m_x $                              & $ \overline{m}_x $          &  $\sigma_d $                        & $ \overline{\sigma}_d  $  &  $ \sigma_d' $                   &  $ \overline{\sigma}_d' $ & $ m_y $                             &  $ \overline{m}_y $             &  $ e $                                 & $\overline{e} $                   &  $C_2$                            & $\overline{C}_2 $          & $C_4 $                                 & $\overline{C}_4 $           &  $ C_4^{-1} $                      & $\overline{C}_4^{-1}$ \\ 
\hline
$ \overline{m}_x $             &  $ \overline{m}_x $            & $ m_x $                             &  $ \overline{\sigma}_d $    & $  \sigma_d  $                    & $ \overline{\sigma}_d' $ &  $ \sigma_d'$                     & $\overline{m}_y $           & $ m_y $                               & $ \overline{e} $                & $ e $                                     & $ \overline{C}_2 $         & $ C_2 $                           & $\overline{C}_4 $               & $ C_4 $                             & $ \overline{C}_4^{-1} $    & $ C_4^{-1}$ \\
\hline
$ m_y $                               &  $ m_y $                              & $ \overline{m}_y $           &  $  \sigma_d' $                     & $ \overline{\sigma}_d'  $  &  $ \sigma_d $                   &  $ \overline{\sigma}_d $  &  $m_x$                              & $\overline{m}_x $             & $C_2 $                              & $ \overline{C}_2 $              & $ e $                                & $ \overline{e} $              & $C_4^{-1} $                         & $ \overline{C}_4^{-1}$   & $ C_4 $                               & $ \overline{C}_4 $ \\
\hline
$ \overline{m}_y $             &  $ \overline{m}_y $            &  $m_y$                             &  $  \overline{\sigma}_d' $   & $ \sigma_d'   $                    & $ \overline{\sigma}_d  $ & $ \sigma_d $                    &  $\overline{m}_x $           & $ m_x $                              & $\overline{C}_2 $            &  $C_2 $                                & $\overline{e} $               &  $ e $                                & $ \overline{C}_4^{-1}$      & $ C_4^{-1} $                    & $ \overline{C}_4 $              & $ C_4 $ \\
\hline
$ \sigma_d $                       &  $ \sigma_d $                     &  $ \overline{\sigma}_d $& $  m_y  $                              & $  \overline{m}_y  $            &  $m_x$                              & $ \overline{m}_x  $         &  $\sigma_d' $                    & $ \overline{\sigma}_d' $  &  $C_4^{-1}$                      &  $\overline{C}_4^{-1} $     & $C_4 $                             & $\overline{C}_4 $          & $ e $                                     & $ \overline{e} $               & $ C_2 $                                 & $\overline{C}_2$ \\
\hline
$ \overline{\sigma}_d $     &  $ \overline{\sigma}_d $  & $ \sigma_d $                   &  $ \overline{m}_y  $            & $ m_y$                                 &  $\overline{m}_x  $          & $ m_x $                             & $\overline{\sigma}_d' $  & $ \sigma_d' $                     &  $ \overline{C}_4^{-1}$  &  $ C_4^{-1} $                       & $\overline{C}_4 $          & $ C_4 $                           & $ \overline{e} $                   &  $ e $                                & $ \overline{C}_2  $              & $ C_2 $\\
\hline
$ \sigma_d' $                      &   $ \sigma_d' $                    & $\overline{\sigma}_d' $&  $ m_x  $                               & $ \overline{m}_x$              &  $m_y$                               & $\overline{m}_y$            &  $ \sigma_d $                   & $ \overline{\sigma}_d $   & $ C_4 $                             & $ \overline{C}_4^{-1} $     &  $ C_4^{-1} $                   & $ \overline{C}_4^{-1} $& $C_2$                                  & $ \overline{C}_2 $         &  $ e $                                      & $ \overline{e} $\\
\hline
$ \overline{\sigma}_d' $    &  $ \overline{\sigma}_d' $  &  $ \sigma_d' $                 &  $ \overline{m}_x $             & $ m_x  $                              &  $\overline{m}_y $            &  $m_y $                            &  $\overline{\sigma}_d  $ & $ \sigma_d $                     & $ \overline{C}_4 $          &  $ C_4 $                                &  $\overline{C}_4^{-1} $ & $ C_4^{-1} $                   & $\overline{C}_2$               & $C_2 $                             & $\overline{e}$                      & $ e $ \\
\hline
\end{tabular}
\end{center}
\label{d4h_mult}
\end{sidewaystable}

\subsection{Character table of unitary subgroup}

Table~\ref{d4h_char} shows the character table for the $D_{4h}$ point group along with the result of performing the Dimmock-Wheeler test.  
Only the bottom five irreducible representations are physically applicable in the present
scenario since the character must change sign between an operator and its overbarred counterpart. 
Interestingly, at the M point one expects both Type (a) and Type (c) corepresentations, and the Type (a) corepresentation is ``classically'' doubly degenerate.

\begin{table}[ht]
\caption{Character table of the unitary subgroup of $\mathbf{k} = (\boldsymbol{x}+\boldsymbol{y})\frac{\pi}{\Lambda}$ which is isomorphic to $D_{4h} (4/mmm)$.}
\begin{center}
\begin{ruledtabular}
\begin{tabular}{c|c c c c c c c c c c | c c}
$D_{4h}$ & $ e $ & $ C_4,C_4^{-1} $ & $ C_2 $ & $ \overline{m}_{x}, \overline{m}_{y} $ & $ \overline{\sigma}_d, \overline{\sigma}_d' $ & $ \overline{e} $ & 
                      $ \overline{C}_4,\overline{C}_4^{-1} $ & $ \overline{C}_2 $ & $ m_x, m_y $ & $\sigma_d, \sigma_d' $ & $\alpha$ & Correp. Type \\
\hline
   $ A_{1g}   $   & 1 &  1 &  1   &  1 &   1  &  1  &  1  &  1  &  1  &  1  &    NA      &     NA \\
   $ A_{2g}   $   & 1 &  1 &  1   & -1 &  -1 &  1  &  1  &  1  & -1  & -1  &    NA      &     NA \\
   $ B_{1g}   $   & 1 & -1 &  1   &  1 &  -1 &  1  & -1  &  1  &  1  & -1  &    NA      &     NA \\
   $ B_{2g}   $   & 1 & -1 &  1   & -1 &   1 &  1  & -1  & -1  & -1  &  1  &    NA      &     NA \\
   $ E_{g}     $   &  2 & 0 & -2   &  0  &   0 &  2  &  0  & -2  &  0  &  0  &    NA      &     NA \\
   $ A_{1u}   $   & 1 &  1 &  1   &   1 &   1 & -1  & -1 & -1  & -1  & -1  &    0         &     (c)\\
   $ A_{2u}   $   & 1 &  1 &  1   &  -1 &  -1 & -1 & -1  & -1  &  1  &  1  &    0         &     (c)\\
   $ B_{1u}   $   & 1 & -1 &  1   &   1 &  -1 & -1 &  1  &  -1 & -1  &  1  &    0         &     (c)\\
   $ B_{2u}   $   & 1 & -1 &  1   &  -1 &  1  & -1 &  1  &  -1 &  1  & -1  &    0         &     (c)\\
   $ E_{u}     $   & 2  &  0 & -2  &    0 &  0  & -2 &  0  &   2 &  0  &  0   &   16       &     (a)\\
\end{tabular}
\end{ruledtabular}
\end{center}
\label{d4h_char}
\end{table}

\subsection{Construction of corepresentations}

For the Type (c) corepresentations, we use 

$A = \xi = S\mathcal{T}$ \\
$S = \{E|2m \Lambda + \Lambda/2, 2n\Lambda + \Lambda/2 \}$\\
$S^{-1} = \{E|-2m\Lambda - \Lambda/2, -2n\Lambda - \Lambda/2 \}$\\
Calculations needed for Supplemental Material Eq.~\ref{eqcorepU} are provided in Table~\ref{tabMuelements} for the unitary elements.
Calculations needed for Supplemental Material Eq.~\ref{eqcorepAU} are provided in Table~\ref{tabMauelements} for the antiunitary elements. \\

\begin{table}[ht]
\caption{Determination of Type (c) corepresentations for unitary elements. }
\begin{center}
\begin{tabular}{c | c | c}
$R$                                      &      $RS$                                                                                                                                       &  $S^{-1}RS$ \\
\hline
$e$                                      &      $ \{E | 2m\Lambda + \Lambda/2, 2n\Lambda + \Lambda/2 \}$                                      & $\{E | 0,0 \} = e$   \\
\hline
$C_4$                                 &      $ \{C_4 | -2m\Lambda - \Lambda/2, 2n\Lambda + \Lambda/2 \}$                                & $\{C_4 | -4m\Lambda-\Lambda, 0 \} = \overline{C}_4 $   \\
\hline
$C_4^{-1} $                       &      $ \{C_4^{-1} | 2m\Lambda + \Lambda/2, -2n\Lambda - \Lambda/2 \}$                         & $\{C_4^{-1} | 0, -4n\Lambda-\Lambda \} = \overline{C}_4^{-1} $   \\
\hline
$C_2$                                &      $ \{C_2 | -2m\Lambda - \Lambda/2, -2n\Lambda - \Lambda/2 \} $                                & $\{C_2 | -2m\Lambda - \Lambda, -2n\Lambda - \Lambda \} = C_2 $   \\
\hline
$\overline{m}_{x} $          &      $ \{\overline{\sigma}_x | -2m\Lambda - \Lambda/2, 2n\Lambda + \Lambda/2 \}$      &  $\{\overline{\sigma}_x | -4m\Lambda-\Lambda, 0 \} = m_x $\\
\hline
$\overline{m}_{y}$           &      $ \{\overline{\sigma}_y | 2m\Lambda + \Lambda/2, -2n\Lambda - \Lambda/2 \}$      &  $\{\overline{\sigma}_y | 0, -4n\Lambda-\Lambda  \} = m_y $ \\
\hline
$\overline{\sigma}_d$     &      $ \{\overline{\sigma}_d | 2m\Lambda + \Lambda/2, 2n\Lambda + \Lambda/2 \}$      &  $\{\overline{\sigma}_d | 0, 0 \} = \overline{\sigma}_d $\\
\hline
$\overline{\sigma}_d'$    &      $ \{\overline{\sigma}_d' | -2m\Lambda - \Lambda/2, -2n\Lambda - \Lambda/2 \}$      &  $\{\overline{\sigma}_d' | -4m\Lambda - \Lambda, -4n\Lambda - \Lambda  \} = \overline{\sigma}_d'$ \\
\hline
$\overline{e}$                   &      $ \{\overline{e} | 2m\Lambda + \Lambda/2, 2n\Lambda +\Lambda/2 \}$                       & $\{\overline{e} | 0,0 \} = \overline{e}$   \\
\hline
$\overline{C}_4$              &      $ \{\overline{C}_4 | -2m\Lambda - \Lambda/2, 2n\Lambda + \Lambda/2 \}$                & $\{\overline{C}_4 | -4m\Lambda-\Lambda, 0 \} = C_4 $   \\
\hline
$\overline{C}_4^{-1} $     &      $ \{\overline{C}_4^{-1} | 2m\Lambda + \Lambda/2, -2n\Lambda - \Lambda/2 \}$       & $\{\overline{C}_4^{-1} | 0, -4n\Lambda - \Lambda \} = C_4^{-1}$   \\
\hline
$\overline{C}_2$              &      $ \{\overline{C}_2 | -2m\Lambda - \Lambda/2, -2n\Lambda - \Lambda/2 \} $               & $\{\overline{C}_2 | -4m\Lambda - \Lambda, -4n\Lambda - \Lambda \} = \overline{C}_2 $   \\
\hline
$m_{x} $                             &      $ \{\sigma_x | -2m\Lambda - \Lambda/2, 2n\Lambda + \Lambda/2 \}$                          &  $\{\sigma_x | -4m\Lambda - \Lambda, 0 \} = \overline{m}_x$\\
\hline
$m_{y}$                              &      $ \{\sigma_y | 2m\Lambda + \Lambda/2, -2n\Lambda - \Lambda/2 \}$                          &  $\{\sigma_y | 0, -2n\Lambda - \Lambda  \} = \overline{m}_y$ \\
\hline
$\sigma_d$                        &      $ \{\sigma_d | 2m\Lambda + \Lambda/2, 2n\Lambda + \Lambda/2 \}$                          &  $\{\sigma_d | 0, 0 \} = \sigma_d $\\
\hline
$\sigma_d'$                       &      $ \{\sigma_d' | -2m\Lambda - \Lambda/2, -2n\Lambda - \Lambda/2 \}$                         &  $\{\sigma_d' | -4m\Lambda - \Lambda, -4n\Lambda - \Lambda \} = \sigma_d'$ \\
\end{tabular}
\end{center}
\label{tabMuelements}
\end{table}

\begin{table}[ht]
\caption{Determination of Type (c) corepresentations for antiunitary elements.  }
\begin{center}
\begin{tabular}{c | c | c }
$B$                                           &      $BA                                 $  & $ A^{-1}B                       $  \\
\hline
$ \xi $                                        &      $ e                                   $  & $ e                                  $   \\
\hline
$ \overline{\xi} $                      &      $ \overline{e}                $   & $ \overline{e}               $   \\
\hline
$\gamma $                               &      $  \overline{C}_4         $   &  $ C_4                            $   \\
\hline
$ \overline{\gamma} $            &      $  C_4                           $   &  $ \overline{C}_4          $   \\
\hline
$\gamma^{-1}$                        &      $ \overline{C}_4^{-1} $   & $ C_4^{-1}                     $   \\
\hline
$\overline{\gamma}_4^{-1}$ &      $ C_4^{-1}                   $   & $  \overline{C}_4^{-1}  $   \\
\hline
$ \zeta $                                    &      $  C_2                           $   &  $ C_2                            $ \\
\hline
$ \overline{\zeta} $                  &      $  \overline{C}_2        $   &  $ \overline{C}_2          $ \\
\hline
$ \mu_x $                                  &      $ \overline{m}_x         $   &  $ m_x                            $\\
\hline
$ \overline{\mu}_x $               &      $  m_x                           $   &  $ \overline{m}_x          $ \\
\hline
$ \mu_y $                                  &      $ \overline{m}_y         $   &  $ m_y                            $  \\
\hline
$ \overline{\mu}_y $                &      $ m_y                           $   &  $ \overline{m}_y          $  \\
\hline
$ \mu_d $                                  &      $ \sigma_d                   $  &   $ \sigma_d                  $ \\
\hline
$ \overline{\mu}_d $                &      $ \overline{\sigma}_d $ &  $ \overline{\sigma}_d $  \\
\hline
$ \mu_d' $                                 &      $ \sigma_d'                   $ &  $ \sigma_d'                    $ \\
\hline
$\overline{\mu}_d' $                &      $ \overline{\sigma}_d' $ &  $ \overline{\sigma}_d'  $ \\
\end{tabular}
\end{center}
\label{tabMauelements}
\end{table}

For the Type (a) corepresentations, we again use $A = \xi$, and the calculations are provided in the following.

$\mathbf{\Gamma}_i(e \xi )                                  = \mathbf{\Gamma}_i(\xi)                                         = \boldsymbol{\beta} \mathbf{\Delta}_i(e)$  \\
$\mathbf{\Gamma}_i(C_4 \xi)                              = \mathbf{\Gamma}_i(\overline{\gamma})                = \boldsymbol{\beta} \mathbf{\Delta}_i(C_4)$  \\
$\mathbf{\Gamma}_i(C_4^{-1} \xi)                      = \mathbf{\Gamma}_i(\overline{\gamma}^{-1})          = \boldsymbol{\beta} \mathbf{\Delta}_i(C_4^{-1})$  \\
$\mathbf{\Gamma}_i(C_2 \xi)                              = \mathbf{\Gamma}_i(\zeta)                                      = \boldsymbol{\beta} \mathbf{\Delta}_i(C_2)$  \\

$\mathbf{\Gamma}_i(\overline{m}_x \xi)             = \mathbf{\Gamma}_i(\mu_x)                                    = \boldsymbol{\beta} \mathbf{\Delta}_i(\overline{m}_x)$  \\
$\mathbf{\Gamma}_i(\overline{m}_y \xi)             = \mathbf{\Gamma}_i(\mu_u)                                   = \boldsymbol{\beta} \mathbf{\Delta}_i(\overline{m}_y)$  \\
$\mathbf{\Gamma}_i(\overline{\sigma}_d \xi)    = \mathbf{\Gamma}_i(\overline{\mu}_d)                    = \boldsymbol{\beta} \mathbf{\Delta}_i(\overline{\sigma}_d)$  \\
$\mathbf{\Gamma}_i(\overline{\sigma}_d' \xi)   = \mathbf{\Gamma}_i(\overline{\mu}_d')                    = \boldsymbol{\beta} \mathbf{\Delta}_i(\overline{\sigma}_d')$  \\

$\mathbf{\Gamma}_i(\overline{e} \xi )                  = \mathbf{\Gamma}_i(\overline{\xi})                       = \boldsymbol{\beta} \mathbf{\Delta}_i(\overline{e})$  \\
$\mathbf{\Gamma}_i(\overline{C}_4 \xi)              = \mathbf{\Gamma}_i(\gamma)                              =  \boldsymbol{\beta} \mathbf{\Delta}_i(\overline{C}_4)$  \\
$\mathbf{\Gamma}_i(\overline{C}_4^{-1} \xi)      = \mathbf{\Gamma}_i(\gamma^{-1})                        = \boldsymbol{\beta} \mathbf{\Delta}_i(\overline{C}_4^{-1})$  \\
$\mathbf{\Gamma}_i(\overline{C}_2 \xi)              = \mathbf{\Gamma}_i(\overline{\zeta})                   =  \boldsymbol{\beta} \mathbf{\Delta}_i(\overline{C}_2)$  \\

$\mathbf{\Gamma}_i(m_x \xi)             = \mathbf{\Gamma}_i(\overline{\mu}_x)                                 = \boldsymbol{\beta} \mathbf{\Delta}_i(m_x)$  \\
$\mathbf{\Gamma}_i(m_y \xi)             = \mathbf{\Gamma}_i(\overline{\mu}_u)                                 = \boldsymbol{\beta} \mathbf{\Delta}_i(m_y)$  \\
$\mathbf{\Gamma}_i(\sigma_d \xi)    = \mathbf{\Gamma}_i(\mu_d)                                                  = \boldsymbol{\beta} \mathbf{\Delta}_i(\sigma_d)$  \\
$\mathbf{\Gamma}_i(\sigma_d' \xi)   = \mathbf{\Gamma}_i(\mu_d')                                                  = \boldsymbol{\beta} \mathbf{\Delta}_i(\sigma_d')$  \\

where $\boldsymbol{\beta}$ is again defined by $\boldsymbol{\beta} \boldsymbol{\beta}^* = {\bf\Delta}_i(E)$.
The full corepresentation is provided in Tables~\ref{corepMU1a}-\ref{corepMAU1b}.

Corepresentations 1 and 4 and corepresentations 2 and 3 are equivalent through the unitary transformation 
\[ U= 
 \begin{pmatrix}
          0  & c    \\
          c*  & 0  \\
 \end{pmatrix}
\]
where $c$ is a complex constant of unity modulus.

In the bandstructure shown in Fig. 2 of the main text, the corepresentations labels that include a superscript $u$ are to be transformed using the same unitary matrix discussed
at the $\Gamma$ point
\[ \boldsymbol{u} = 
 \begin{pmatrix}
          1  & 0    \\
          0  & -1  \\
 \end{pmatrix} 
\]
The corepresentation label with a superscript $\beta$ indicates that the antiunitary matrices should be multplied by 
\[ \boldsymbol{\beta} = 
 \begin{pmatrix}
         -1   &  0   \\
          0   &  1   \\
 \end{pmatrix}.
\]

\begin{sidewaystable}[]
\caption{ Corepresentations of the group of $\mathbf{k} = (\boldsymbol{x} + \boldsymbol{y})\frac{\pi}{\Lambda}$ for the unitary symmetry operators.  (1/2) }
\begin{center}
\begin{ruledtabular}
\begin{tabular}{c c | c c c c c c c c c c }
Type  &  $D_{4h} (4/mmm)$ & $ e $ & $ C_4,C_4^{-1} $ & $ C_2 $ & $ \overline{m}_x, \overline{m}_y $ & $ \overline{\sigma}_d, \overline{\sigma}_d' $   \\
\hline
  (c)  &  $ A_{1u},        \mathbf{\Gamma}_1 $       &  
  
$\left( \begin{tabular}{cc}
         1 & 0  \\
         0 & 1  \\
        \end{tabular}
        \right)$
&
$\left( \begin{tabular}{cc}
         1 & 0  \\
         0 & -1  \\
        \end{tabular}
        \right)$
& 
$\left( \begin{tabular}{cc}
         1 & 0  \\
         0 & 1  \\
        \end{tabular}
        \right)$
&
$\left( \begin{tabular}{cc}
         1 &  0  \\
         0 & -1  \\
        \end{tabular}
        \right)$
&
$\left( \begin{tabular}{cc}
         1 &   0  \\
         0 &   1  \\
        \end{tabular}
        \right)$
\\

  (c)  &  $ A_{2u},  \mathbf{\Gamma}_2 $       & 
  
$\left( \begin{tabular}{cc}
         1 & 0  \\
         0 & 1  \\
        \end{tabular}
        \right)$
&
$\left( \begin{tabular}{cc}
          1 & 0  \\
         0 & -1  \\
        \end{tabular}
        \right)$
& 
$\left( \begin{tabular}{cc}
         1 & 0  \\
         0 & 1  \\
        \end{tabular}
        \right)$
&
$\left( \begin{tabular}{cc}
         -1 & 0  \\
         0 & 1  \\
        \end{tabular}
        \right)$
&
$\left( \begin{tabular}{cc}
         -1 & 0  \\
         0 & -1  \\
        \end{tabular}
        \right)$
\\  

 (c)  &  $ B_{1u},  \mathbf{\Gamma}_3 $       & 
  
  $\left( \begin{tabular}{cc}
         1 & 0  \\
         0 & 1  \\
        \end{tabular}
        \right)$
&
$\left( \begin{tabular}{cc}
         -1 & 0  \\
         0 &  1  \\
        \end{tabular}
        \right)$
& 
$\left( \begin{tabular}{cc}
         1 & 0  \\
         0 & 1  \\
        \end{tabular}
        \right)$
&
$\left( \begin{tabular}{cc}
         1 &  0  \\
         0 & -1  \\
        \end{tabular}
        \right)$
&
$\left( \begin{tabular}{cc}
         -1 & 0  \\
          0 & -1  \\
        \end{tabular}
        \right)$
\\

  (c)  &  $ B_{2u},  \mathbf{\Gamma}_4 $       & 

$\left( \begin{tabular}{cc}
         1 & 0  \\
         0 & 1  \\
        \end{tabular}
        \right)$
&
$\left( \begin{tabular}{cc}
         -1 & 0  \\
         0 & 1  \\
        \end{tabular}
        \right)$
& 
$\left( \begin{tabular}{cc}
         1 & 0  \\
         0 & 1  \\
        \end{tabular}
        \right)$
&
$\left( \begin{tabular}{cc}
         -1 & 0  \\
         0 &  1  \\
        \end{tabular}
        \right)$
&
$\left( \begin{tabular}{cc}
         1 & 0  \\
         0 & 1  \\
        \end{tabular}
        \right)$
\\

  (a)  &  $ E_{u},  \mathbf{\Gamma}_5 $       & 

$\left( \begin{tabular}{cc}
         1 & 0  \\
         0 & 1  \\
        \end{tabular}
        \right)$
&
$\left( \begin{tabular}{cc}
         0 & -1  \\
         1 & 0  \\
        \end{tabular}
        \right)$,
$\left( \begin{tabular}{cc}
         0 &  1  \\
        -1 &  0  \\
        \end{tabular}
        \right)$
&
$\left( \begin{tabular}{cc}
         -1 & 0  \\
          0 & -1  \\
        \end{tabular}
        \right)$
&
$\left( \begin{tabular}{cc}
         0 & -1  \\
         -1 & 0  \\
        \end{tabular}
        \right)$,
$\left( \begin{tabular}{cc}
          0 & 1  \\
         1 &  0  \\
        \end{tabular}
        \right)$
& 
$\left( \begin{tabular}{cc}
         1 & 0  \\
         0 & -1  \\
        \end{tabular}
        \right)$,
$\left( \begin{tabular}{cc}
         -1 & 0  \\
         0 &  1  \\
        \end{tabular}
        \right)$
\\
\end{tabular}
\end{ruledtabular}
\end{center}
\label{corepMU1a}
\end{sidewaystable}

\begin{sidewaystable}[]
\caption{ Corepresentations of the group of $\mathbf{k} = (\boldsymbol{x} + \boldsymbol{y})\frac{\pi}{\Lambda}$ for the unitary symmetry operators.  (2/2)  }
\begin{center}
\begin{ruledtabular}
\begin{tabular}{c c | c c c c c c c c c c }
Type  &  $D_{4h} (4/mmm)$ & $ \overline{e} $ & $ \overline{C}_4, \overline{C}_4^{-1}$ & $\overline{C}_2 $ & $ m_x, m_y $ & $ \sigma_d, \sigma_d' $ \\
\hline
  (c)  &  $ A_{1u},        \mathbf{\Gamma}_1 $       &
  
$\left( \begin{tabular}{cc}
        -1 &  0  \\
         0 & -1  \\
        \end{tabular}
        \right)$
& 
$\left( \begin{tabular}{cc}
        -1 &  0  \\
         0 &  1  \\
        \end{tabular}
        \right)$
&
$\left( \begin{tabular}{cc}
        -1 &  0  \\
         0 & -1  \\
        \end{tabular}
        \right)$
&
$\left( \begin{tabular}{cc}
        -1 &  0  \\
         0 &  1  \\
        \end{tabular}
        \right)$
&
$\left( \begin{tabular}{cc}
         -1 &  0  \\
          0 & -1  \\
        \end{tabular}
        \right)$
\\

  (c)  &  $ A_{2u},  \mathbf{\Gamma}_2 $       & 
  
$\left( \begin{tabular}{cc}
         -1 &  0  \\
          0 & -1  \\
        \end{tabular}
        \right)$
& 
$\left( \begin{tabular}{cc}
         -1 & 0  \\
          0 & 1  \\
        \end{tabular}
        \right)$
&
$\left( \begin{tabular}{cc}
         -1 & 0  \\
          0 & -1  \\
        \end{tabular}
        \right)$
&
$\left( \begin{tabular}{cc}
          1 & 0  \\
          0 & -1  \\
        \end{tabular}
        \right)$
&
$\left( \begin{tabular}{cc}
          1 & 0  \\
          0 & 1  \\
        \end{tabular}
        \right)$
\\  

 (c)  &  $ B_{1u},  \mathbf{\Gamma}_3 $       & 
  
$\left( \begin{tabular}{cc}
         -1 &  0  \\
         0 & -1  \\
        \end{tabular}
        \right)$
& 
$\left( \begin{tabular}{cc}
          1 & 0  \\
          0 & -1  \\
        \end{tabular}
        \right)$
&
$\left( \begin{tabular}{cc}
         -1 & 0  \\
         0 & -1  \\
        \end{tabular}
        \right)$
& 
$\left( \begin{tabular}{cc}
          -1 & 0  \\
          0 &  1  \\
        \end{tabular}
        \right)$
&
$\left( \begin{tabular}{cc}
         1 & 0  \\
         0 & 1  \\
        \end{tabular}
        \right)$
\\

  (c)  &  $ B_{2u},  \mathbf{\Gamma}_4 $       & 

$\left( \begin{tabular}{cc}
         -1 & 0  \\
         0 & -1  \\
        \end{tabular}
        \right)$
& 
$\left( \begin{tabular}{cc}
         1 & 0  \\
         0 & -1  \\
        \end{tabular}
        \right)$
&
$\left( \begin{tabular}{cc}
         -1 & 0  \\
         0 & -1  \\
        \end{tabular}
        \right)$
& 
$\left( \begin{tabular}{cc}
         1 & 0  \\
         0 & -1  \\
        \end{tabular}
        \right)$
&
$\left( \begin{tabular}{cc}
         -1 & 0  \\
         0 & -1  \\
        \end{tabular}
        \right)$
\\

  (a)  &  $ E_{u},  \mathbf{\Gamma}_5 $       & 
  
$\left( \begin{tabular}{cc}
         -1 & 0  \\
         0 & -1  \\
        \end{tabular}
        \right)$
&
$\left( \begin{tabular}{cc}
         0 & 1  \\
        -1 & 0  \\
        \end{tabular}
        \right)$,
$\left( \begin{tabular}{cc}
         0 & -1  \\
         1 &  0  \\
        \end{tabular}
        \right)$
&
$\left( \begin{tabular}{cc}
         1 & 0  \\
          0 & 1  \\
        \end{tabular}
        \right)$
&
$\left( \begin{tabular}{cc}
         0 &  1  \\
          1 & 0  \\
        \end{tabular}
        \right)$,
$\left( \begin{tabular}{cc}
          0 & -1  \\
         -1 &  0  \\
        \end{tabular}
        \right)$
& 
$\left( \begin{tabular}{cc}
         -1 & 0  \\
         0 & 1  \\
        \end{tabular}
        \right)$,
$\left( \begin{tabular}{cc}
         1 & 0  \\
         0 &  -1  \\
        \end{tabular}
        \right)$
\\
\end{tabular}
\end{ruledtabular}
\end{center}
\label{corepMU1b}
\end{sidewaystable}

\begin{sidewaystable}[]
\caption{ Corepresentations of the group of $\mathbf{k} = (\boldsymbol{x} + \boldsymbol{y})\frac{\pi}{\Lambda}$ for the antiunitary symmetry operators. (1/2)  }
\begin{center}
\begin{ruledtabular}
\begin{tabular}{c c | c c c c c c c c c c }
Type  &  $D_{4h} (4/mmm)$ & $ \xi $ & $ \gamma, \gamma^{-1} $ & $ \zeta $ & $ \overline{\mu}_x, \overline{\mu}_y $ & $ \overline{\mu}_d, \overline{\mu}_d' $ \\
\hline
  (c)  &  $ A_{1u},        \mathbf{\Gamma}_1 $       &  
  
$\left( \begin{tabular}{cc}
         0 & 1  \\
         1 & 0  \\
        \end{tabular}
        \right)$
&
$\left( \begin{tabular}{cc}
         0 & 1  \\
        -1 & 0  \\
        \end{tabular}
        \right)$
& 
$\left( \begin{tabular}{cc}
         0 & 1  \\
         1 & 0  \\
        \end{tabular}
        \right)$
&
$\left( \begin{tabular}{cc}
         0 &  1  \\
        -1 & 0  \\
        \end{tabular}
        \right)$
&
$\left( \begin{tabular}{cc}
         0 &  1  \\
         1 &  0  \\
        \end{tabular}
        \right)$
\\

  (c)  &  $ A_{2u},  \mathbf{\Gamma}_2 $       & 
  
$\left( \begin{tabular}{cc}
         0 & 1  \\
         1 & 0  \\
        \end{tabular}
        \right)$
&
$\left( \begin{tabular}{cc}
          0 & 1  \\
         -1 & 0  \\
        \end{tabular}
        \right)$
& 
$\left( \begin{tabular}{cc}
         0 & 1  \\
         1 & 0  \\
        \end{tabular}
        \right)$
&
$\left( \begin{tabular}{cc}
          0 & -1  \\
          1 &  0  \\
        \end{tabular}
        \right)$
&
$\left( \begin{tabular}{cc}
          0 & -1  \\
         -1 & 0  \\
        \end{tabular}
        \right)$
\\  

 (c)  &  $ B_{1u},  \mathbf{\Gamma}_3 $       & 
  
  $\left( \begin{tabular}{cc}
         0 & 1  \\
         1 & 0  \\
        \end{tabular}
        \right)$
&
$\left( \begin{tabular}{cc}
         0 & -1  \\
         1 &  0  \\
        \end{tabular}
        \right)$
& 
$\left( \begin{tabular}{cc}
         0 & 1  \\
         1 & 0  \\
        \end{tabular}
        \right)$
&
$\left( \begin{tabular}{cc}
         0 &  1  \\
        -1 &  0  \\
        \end{tabular}
        \right)$
&
$\left( \begin{tabular}{cc}
         0 &  -1  \\
        -1 &   0  \\
        \end{tabular}
        \right)$
\\

  (c)  &  $ B_{2u},  \mathbf{\Gamma}_4 $       & 

$\left( \begin{tabular}{cc}
         0 & 1  \\
         1 & 0  \\
        \end{tabular}
        \right)$
&
$\left( \begin{tabular}{cc}
         0 & -1  \\
         1 &  0  \\
        \end{tabular}
        \right)$
& 
$\left( \begin{tabular}{cc}
         0 & 1  \\
         1 & 0  \\
        \end{tabular}
        \right)$
&
$\left( \begin{tabular}{cc}
         0 & -1  \\
         1 &  0  \\
        \end{tabular}
        \right)$
&
$\left( \begin{tabular}{cc}
         0 & 1  \\
         1 & 0  \\
        \end{tabular}
        \right)$
\\

  (a)  &  $ E_{u},  \mathbf{\Gamma}_5 $       & 
  
$\left( \begin{tabular}{cc}
         1 & 0  \\
         0 & 1  \\
        \end{tabular}
        \right)$
&
$\left( \begin{tabular}{cc}
         0 &   1  \\
         -1 & 0  \\
        \end{tabular}
        \right)$,
$\left( \begin{tabular}{cc}
         0 & -1  \\
         1 &  0  \\
        \end{tabular}
        \right)$
&
$\left( \begin{tabular}{cc}
         -1 & 0  \\
          0 & -1  \\
        \end{tabular}
        \right)$
&
$\left( \begin{tabular}{cc}
         0 &  1  \\
          1 & 0  \\
        \end{tabular}
        \right)$,
$\left( \begin{tabular}{cc}
          0 & -1  \\
         -1 &  0  \\
        \end{tabular}
        \right)$
&
$\left( \begin{tabular}{cc}
         1 & 0  \\
         0 & -1  \\
        \end{tabular}
        \right)$,
$\left( \begin{tabular}{cc}
         -1 & 0  \\
         0 &  1  \\
        \end{tabular}
        \right)$

\end{tabular}
\end{ruledtabular}
\end{center}
\label{corepMAU1a}
\end{sidewaystable}

\begin{sidewaystable}[]
\caption{ Corepresentations of the group of $\mathbf{k} = (\boldsymbol{x} + \boldsymbol{y})\frac{\pi}{\Lambda}$ for the antiunitary symmetry operators.  (2/2)  }
\begin{center}
\begin{ruledtabular}
\begin{tabular}{c c | c c c c c c c c c c }
Type  &  $D_{4h} (4/mmm)$ &  $ \overline{\xi} $ & $ \overline{\gamma}, \overline{\gamma}^{-1}$ & $\overline{\zeta} $ & $ \mu_x, \mu_y $ & $ \mu_d, \mu_d' $ \\
\hline
  (c)  &  $ A_{1u},        \mathbf{\Gamma}_1 $       &  
  
$\left( \begin{tabular}{cc}
         0 &  -1  \\
        -1 &   0  \\
        \end{tabular}
        \right)$
& 
$\left( \begin{tabular}{cc}
         0 & -1  \\
         1 &  0  \\
        \end{tabular}
        \right)$
&
$\left( \begin{tabular}{cc}
          0 &  -1  \\
         -1 &  0  \\
        \end{tabular}
        \right)$
&
$\left( \begin{tabular}{cc}
         0 & -1  \\
         1 &  0  \\
        \end{tabular}
        \right)$
&
$\left( \begin{tabular}{cc}
          0 &  -1  \\
         -1 &  0  \\
        \end{tabular}
        \right)$
\\

  (c)  &  $ A_{2u},  \mathbf{\Gamma}_2 $       & 
  
$\left( \begin{tabular}{cc}
         0 &  -1  \\
        -1 &   0  \\
        \end{tabular}
        \right)$
& 
$\left( \begin{tabular}{cc}
          0 & -1  \\
          1 &  0  \\
        \end{tabular}
        \right)$
&
$\left( \begin{tabular}{cc}
         0 &  -1  \\
        -1 &   0  \\
        \end{tabular}
        \right)$
&
$\left( \begin{tabular}{cc}
          0 &  1  \\
         -1 & 0  \\
        \end{tabular}
        \right)$
&
$\left( \begin{tabular}{cc}
          0 & 1  \\
          1 & 0  \\
        \end{tabular}
        \right)$
\\  

 (c)  &  $ B_{1u},  \mathbf{\Gamma}_3 $       & 
  
$\left( \begin{tabular}{cc}
         0 &  -1  \\
        -1 &   0  \\
        \end{tabular}
        \right)$
& 
$\left( \begin{tabular}{cc}
          0 & 1  \\
         -1 & 0  \\
        \end{tabular}
        \right)$
&
$\left( \begin{tabular}{cc}
         0 &  -1  \\
        -1 &   0  \\
        \end{tabular}
        \right)$
& 
$\left( \begin{tabular}{cc}
          0 & -1  \\
          1 &  0 \\
        \end{tabular}
        \right)$
&
$\left( \begin{tabular}{cc}
         0 & 1  \\
         1 & 0  \\
        \end{tabular}
        \right)$
\\

  (c)  &  $ B_{2u},  \mathbf{\Gamma}_4 $       & 

$\left( \begin{tabular}{cc}
         0 &  -1  \\
        -1 &   0  \\
        \end{tabular}
        \right)$
& 
$\left( \begin{tabular}{cc}
          0 & 1  \\
         -1 & 0  \\
        \end{tabular}
        \right)$
&
$\left( \begin{tabular}{cc}
         0 &  -1  \\
        -1 &   0  \\
        \end{tabular}
        \right)$
& 
$\left( \begin{tabular}{cc}
         0 & 1  \\
        -1 & 0  \\
        \end{tabular}
        \right)$
&
$\left( \begin{tabular}{cc}
          0 & -1  \\
         -1 & 0  \\
        \end{tabular}
        \right)$
\\

  (a)  &  $ E_{u},  \mathbf{\Gamma}_5 $       & 
  
$\left( \begin{tabular}{cc}
         -1 & 0  \\
          0 & -1  \\
        \end{tabular}
        \right)$
&
$\left( \begin{tabular}{cc}
         0 & -1  \\
         1 & 0  \\
        \end{tabular}
        \right)$,
$\left( \begin{tabular}{cc}
         0 &  1  \\
         -1 &  0  \\
        \end{tabular}
        \right)$
&
$\left( \begin{tabular}{cc}
          1 & 0  \\
          0 & 1  \\
        \end{tabular}
        \right)$
&
$\left( \begin{tabular}{cc}
          0 & -1  \\
         -1 & 0  \\
        \end{tabular}
        \right)$,
$\left( \begin{tabular}{cc}
          0 & 1  \\
         1 &  0  \\
        \end{tabular}
        \right)$
&
$\left( \begin{tabular}{cc}
         -1 & 0  \\
         0 & 1  \\
        \end{tabular}
        \right)$,
$\left( \begin{tabular}{cc}
         1 & 0  \\
         0 & -1  \\
        \end{tabular}
        \right)$
\end{tabular}
\end{ruledtabular}
\end{center}
\label{corepMAU1b}
\end{sidewaystable}

\bibliography{IEEEabrv,/Users/adammock1/Desktop/bib/myRef}